\documentclass[10pt,a4paper,aps,prd,twocolumn,showpacs,superscriptaddress]{revtex4-1} 
\usepackage[english]{babel}
\usepackage{url} 
\usepackage{hyperref}
\usepackage{amsmath}
\usepackage{slashed} 
\usepackage{graphicx}

\usepackage{color}
\usepackage{bm} 
\usepackage[capitalize]{cleveref}
\usepackage[normalem]{ulem}

\def\lb{{\mathchar'26\mkern-10mu\lambda}}

\usepackage{multirow}
\usepackage{tabularx}

\usepackage{subcaption}
\begin{document}
\allowdisplaybreaks

\title{Radiation Reaction near the Classical Limit in Aligned Crystals }

\author{C. F. Nielsen}
\affiliation{Department of Physics and Astronomy, Aarhus University, 8000 Aarhus, Denmark}
\author{J. B. Justesen}
\affiliation{Department of Physics and Astronomy, Aarhus University, 8000 Aarhus, Denmark}
\author{A. H. S\o rensen}
\affiliation{Department of Physics and Astronomy, Aarhus University, 8000 Aarhus, Denmark}
\author{U. I. Uggerh\o j}
\affiliation{Department of Physics and Astronomy, Aarhus University, 8000 Aarhus, Denmark}
\author{R. Holtzapple}
\affiliation{Department of Physics, California Polytechnic State University, San Luis Obispo, California 93407, USA}
\collaboration{CERN NA63}
\date{\today}

\begin{abstract}
An accelerated charged particle emits electromagnetic radiation. If the driving force is sufficiently strong, the radiated energy becomes comparable to the kinetic energy of the particle and the back-action of the emitted radiation (radiation reaction) significantly alters the dynamics of the particle. The Landau-Lifshitz (LL) equation has been proposed as the classical equation to describe the dynamics of a charged particle in a strong electromagnetic field when the effects of radiation reaction are taken into account. Hitherto, the experimental problem in validating the LL equation has been to achieve sufficiently strong fields for radiation reaction to be important without quantum effects being prominent. Notwithstanding, here we  provide a quantitative experimental test of the LL equation by measuring the emission spectrum for a wide range of settings for 50 GeV positrons crossing aligned silicon single crystals near the $(110)$ planar channeling regime as well as 40 GeV and 80 GeV electrons traversing aligned diamond single crystals near the $\langle100\rangle$ axial channeling regime. The experimental spectra are in remarkable agreement with predictions based on the LL equation of motion with small quantum corrections for recoil and, in case of electrons, spin and reduced radiation emission, as well as with a more elaborate quantum mechanical model. Our experiment clearly shows the inadequacy of the Lorentz force as the sole agent of force on the particles in the classical limit, due to its absence of radiative energy loss in describing the dynamics of high-energy charged particles in strong electromagnetic fields like those in aligned single crystals.
\end{abstract}

\maketitle
\section{Introduction}
The Lorentz force accounts for the dynamics of charged particles moving in the presence of electromagnetic fields and represents a cornerstone of classical electrodynamics \citep{Jackson_b_1975,Landau_b_2_1975}. This fundamental relation is complemented by Maxwell's equations, which describe the evolution of the electromagnetic field due to charged particles in motion. A well-known consequence of Maxwell's equations is that accelerated charged particles emit electromagnetic radiation \cite{Jackson_b_1975,Landau_b_2_1975}. Under typical experimental conditions the energy radiated by a charged particle is negligible compared to its kinetic energy, such that the radiation can be safely neglected in determining the dynamics of the particle. However, it was realized already at the beginning of the twentieth century \cite{Abraham_b_1905,Lorentz_b_1909,Dirac_1938} that if the particle undergoes large accelerations, the amount of energy radiated becomes comparable to its kinetic energy such that it is essential to include the back-action of the radiation, called the radiation reaction, on the dynamics of the particle \cite{Barut_b_1980,Rohrlich_b_2007}. Such large accelerations, due to strong external electromagnetic fields (called strong-field effects), are achievable for ultrarelativistic particles penetrating single crystals. The reader is referred to \cite{McDo18} for a thorough discussion of the history of the radiation reaction problem, and to \cite{Baie98,Ugge05,Sorensen1996} for an introduction to strong-field effects in crystals.

The equation of motion for a light, charged particle, in a strong external electromagnetic field must take into account the reaction of the radiation on its dynamics. This is done by the Lorentz-Abraham-Dirac (LAD) equation \cite{Abraham_b_1905,Lorentz_b_1909,Dirac_1938}. The LAD equation, however, has peculiar and problematic features. They originate in a dependence of the additional force due to radiation reaction on the time-derivative of the acceleration, which makes the LAD equation structurally ``non-Newtonian''. The presence of the derivative of the acceleration allows for the existence of unphysical (``runaway'') solutions, with the particle acceleration increasing exponentially even if no external field is present. Runaway solutions can be removed by transforming the LAD equation into an integro-differential equation \cite{Rohrlich_b_2007}. However, as a result of this remedy, the particle starts to accelerate before it is acted upon by the external force, which violates the causality principle. 

The Landau-Lifshitz (LL) equation provides an alternative to the controversial LAD equation. It rests on a perturbation expansion which requires the radiation-reaction force to be much smaller than the Lorentz force in the instantaneous rest frame of the charged particle \cite{Landau_b_2_1975}. For an electron moving in an external static electric field $\bf{E}$ and vanishing magnetic field the LL equation reduces to \cite{Landau_b_2_1975}
\begin{equation}
\begin{split}
\frac{d\bf{p}}{dt} = &e{\bf{E}}+\frac{2e^3}{3mc^3}\left[\gamma\left(\bf{v}\cdot{\bm{\nabla}}\right){\bf{E}}+
\frac{e}{mc^2}(\bf{v}\cdot\bf{E})\bf{E}\right.\\
&\left.-\gamma^2\frac{e}{mc^2}
\bf{v}\left(\bf{E}^2-\left(\frac{\bf{v}\cdot\bf{E}}{c}\right)^2\right)\right].
\end{split}
\label{eq:LL}
\end{equation}
Here $e=-\vert e\vert$ is the charge of the electron, ${\bf{p}}=\gamma m {\bf{v}}$ its momentum, $m$ its mass, and $\bf{v}$ its velocity while $\gamma =(1-v^2/c^2)^{-1/2}$ denotes the Lorentz factor. 

The radiation reaction phenomenon, and its relation to the LL equation, have been under active investigation in recent years both theoretically \cite{Murat2019,MuratAndersen2019,Vranic_2014,Blackburn_2014,Tamburini_2014,Li_2014,Heinzl_2015,Yoffe_2015,Capdessus_2015,Vranic_2016,Dinu_2016,Di_Piazza_2017,Harvey_2017,Ridgers_2017,Niel_2017,Niel_2018} and to some extent experimentally \cite{Wistisen_2018,Cole_2018,Poder_2017} (see the recent reviews \cite{Hammond_2010,Di_Piazza_2012,Burton_2014} for previous publications). This paper addresses the validity of the LL equation in characterizing radiation reaction near the classical limit by comparing theoretical simulations based on the LL equation to a combination of new and previously published \cite{cern2017} experimental radiation spectra recorded for high-energy electrons and positrons penetrating aligned single crystals.

The dynamics of the particles, and the emitted radiation, is sensitive to the magnitude of the quantum nonlinearity/strong-field parameter $\chi$ defined as\begin{equation}
             \chi^2=(F_{\mu\nu}u^\nu)^2/E_0^2, \;\;\; E_0=m^2c^3/e\hbar \; ,
 \label{eq:chi}
 \end{equation}
where $F^{\mu\nu}$ is the electromagnetic field tensor, $u^{\nu}$ the four-velocity of the electron (in units of $c$), and $E_0\simeq1.32\times 10^{16}$ V/cm the critical field, see e.g. \cite{Berestetskii_b_1989}. The classical limit is reached for $\chi$ tending to 0, while quantum effects influence the emission spectra already well before $\chi$ approaches unity and are dominant for large values of $\chi$. For emission in a constant field, $\chi$ is the ratio of $\hbar\omega_c$ to the primary energy (up to a factor of 3 depending on the convention for the numerical factor entering the definition of $\omega_c$), where $\omega_c$ is the critical frequency for synchrotron radiation according to classical electrodynamics. For a field that is purely electric ($E$) in the laboratory and transverse to the direction of motion, which is essentially what is experienced by a positron or an electron penetrating a crystal under or near channeling conditions, we have $\chi =\gamma E/E_0$. Note that in this case Eq. (\ref{eq:LL}) simplifies as two terms disappear.

The condition for application of the perturbation approach used to derive the LL equation is that the fields experienced by the radiating electron or positron in its rest frame are small compared with $m^2c^4/e^3$ \cite{Landau_b_2_1975}. This may be expressed as $\delta \ll 1$ where the classical parameter $\delta$, the aforementioned ratio, may conveniently be expressed as $\delta =\chi\alpha$ although neither $\chi$ nor the fine-structure constant $\alpha = e^2/\hbar c \simeq 1/137$ appear in classical physics. For a highly relativistic electron or positron, the radiative damping force is proportional to $\gamma^2$, Eq.\ (\ref{eq:LL}), and for a purely electric field transverse to the direction of motion the ratio of the damping force to the external force is $\eta =\gamma^2r_e\vert e \vert E/mc^2$ (up to a factor of 2/3) where $r_e=e^2/mc^2$ denotes the classical electron radius. Alternatively, $\eta$ may be expressed as $\eta=\gamma\delta$, that is, 
\begin{equation}  
        \eta =\alpha\gamma\chi = \alpha \gamma^2 E/E_0 \; .
         \label{eq:eta}
\end{equation} 
For experimental investigations approaching the classical regime, i.e. for $\chi \ll 1$, it is therefore necessary to have large Lorentz-factors, $\gamma \gg 1$, for the magnitude of the radiation damping force to be appreciable in comparison with the Lorentz force, that is, to achieve a non-negligible $\eta$. As emphasized by Landau and Lifshitz in a footnote \cite{Landau_b_2_1975}, and likely surprising on first account, a large value of the ratio of the damping force to the Lorentz force "does not in any way contradict" the application of the perturbation approach to the derivation of the LL equation.
Since the damping force is longitudinal and the primary force due to the crystal field transverse, their ratio is not Lorentz invariant. After stating the formula for the ratio of the damping force to the external force, Landau and Lifshitz derive an energy $\mathcal{E}_{\text{crit}}$ that the particle cannot exceed after passing through the external field.

In the planning of the experiments the expected percentage of energy lost to radiation is important. It can be estimated as the average radiative damping force times crystal thickness $d_c$ divided by primary energy $\gamma mc^2$. For a transverse electric field this ratio may be expressed as
\begin{equation}
     \Delta = 2d_c \langle \chi^2\rangle /3\gamma a_0 =2d_c \langle \chi^2\rangle \alpha/3\gamma \lb_C \; ,
         \label{eq:LLloss}
\end{equation} 
where $a_0\equiv\hbar^2 /me^2$ is the Bohr radius (of hydrogen) and $\lb_C \equiv\hbar /mc$ the (reduced) Compton wavelength of the electron.

Under our experimental conditions we have $\chi\lesssim0.1$ so $\delta \ll 1$ is clearly fulfilled. 
With $\eta$ attaining values roughly in the range 10--100 the radiation-reaction force dominates the dynamics of the particles while $\chi$ is still sufficiently small that the influence of quantum effects is expected to be moderate. Quantum effects could be avoided by using weaker fields, i.e. smaller $\chi$, but the magnitude of the LL force would then decrease and its effect would be difficult to detect. Our experimental conditions therefore provide an ideal avenue for testing the applicability of the Landau-Lifshitz equation. As a ``standard accelerator'' comparison, the magnetic dipole energy loss according to the Lienard formula takes place with a characteristic distance given as $c\tau=3\chi^{-2}\gamma a_0/2$ ($=d_c/\Delta$) in the high-energy limit, where, 
in this magnetic case, $\chi=\gamma B/B_0$ where $B_0=4.41\times 10^9$ T ($cB_0=E_0$). For a 50 GeV electron in a 2 T field this distance is 6 orders of magnitude larger than the thicknesses of the crystals in our experiments as it equals almost 4 km, a distance over which the particle energy loss has to be replenished by the RF system for the particle to remain in the accelerator.

We report experimental results on the spectrum of photons emitted by positrons and electrons crossing aligned single crystals in the regimes of planar and axial channeling. The dynamics of the charged particles is strongly altered by the self electromagnetic field. Our study presents the first high-statistics quantitative test of the classical LL equation near the classical limit by measuring the single photon emission spectra of electric charges accelerated in strong background fields. Previous tests employing intense laser radiation \cite{Cole_2018,Poder_2017} were based on comparison between the final and the initial energy distribution of ultrarelativistic electrons interacting with a tightly focused terawatt laser, a setup with challenges e.g. in terms of shot-to-shot stability. An advantage of using aligned single crystals is that the electric fields are stationary, stable, and well described, while the fields produced by high-intensity lasers are inherently unstable.
In this experiment 50 GeV positrons cross silicon single crystals in directions close to $(110)$ planes and 40 -- 80 GeV electrons cross diamond single crystals in directions close to the $\langle100\rangle$ axis.

Other applications of crystals to address the radiation reaction process have been investigated theoretically including exploration of the change in angular divergence of electrons and positrons traversing an oriented single crystal close to the channeling regime \cite{Murat1992, Murat1993}.

An attractive phenomenon to investigate using the interaction of electrons with crystalline fields, that is not easily accessible with a laser field, is the Schott term in the classical radiation reaction. The Schott term, which is the first term in square brackets in Eq.\ (\ref{eq:LL}), contains a derivative with respect to position. Since the crystalline fields have a rapid transverse variation, they vary over an {\AA} or less whereas laser wavelengths are of the order microns, the fields from a crystal have a significant advantage to disclose the effect of the Schott term. We have, however, investigated this in the present context and have found in simulations that under our experimental conditions the effect of the Schott term, taking the effect of multiple scattering into account, amounts to at most a few percent and, hence, is too small to detect reliably.

\section{Simulation}
\subsection{Trajectories}
When a charged particle is incident at a small angle to a major crystallographic direction, its motion is in first approximation governed by successive, correlated small-angle collisions with screened target nuclei. Effectively, the trajectory of the particle is determined by the continuum potential obtained by smearing the atomic charges along the axis or the planes with which it is nearly aligned, \cite{Lind65, JUAnotes} and \cite{Sorensen1996, Ugge05}. The continuum potential (energy) varies in the two directions transverse to the axis in the former case, and in the single transverse direction to the planes in the latter. In the axial case, for instance, the continuum potential reads
\begin{equation}
   U(r)\; = \; \frac{1}{d}\int_{-\infty}^{\infty}dzV(r,z) 
               \; ,    \label{eq:contPot}
\end{equation}
where $V$ denotes the potential energy pertaining to the interaction between the projectile and a target atom, $z$ is the coordinate along the atomic row, $r$ the transverse distance to the center of the axis, and $d$ is the average spacing between atoms along it. For a single isolated row of atoms, $U$ has rotational symmetry, Eq.\ (\ref{eq:contPot}). For a true crystal there will be a periodicity in transverse space, $U=U({\bf{r}})$. In the numerical work we use the %so-called 
Doyle-Turner potential which is based on an analytical approximation to relativistic Hartree-Fock atomic potentials. For a single row of atoms and unit projectile charge it reads
\begin{equation}
   U(r)\; = \; \pm\frac{e^2}{a_0}\frac{2a_0^2}{d}
        \sum_{i=1}^4\frac{a_i}{C_i} 
         \, e^{-r^2/C_i} 
               \; ,    \label{eq:DT}
\end{equation}
where the sign reflects that of the incoming charge ($\pm \vert e\vert$),
\begin{equation}
  C_i=C_i(\rho )\equiv b_i/4\pi^2+\rho^2
            \;  ,       \label{eq:DTCi}
\end{equation}
and $\rho$ denotes the two-dimensional root-mean-square thermal displacement of the atom from the equilibrium position. For details and explicit values of the coefficients $a_i$ (\AA ) and $b_i$ (\AA$^2$) see \cite{Doyl68} and \cite{Ande82}, for values of the thermal vibration amplitude see \cite{Nielsen_1980}. The Doyle-Turner potential for a single continuum plane similarly takes the form
\begin{equation}
   U(x)\; = \; \pm 2\pi^{1/2}\frac{e^2}{a_0}a_0^2nd_p
        \sum_{i=1}^4\frac{a_i}{C_i^{1/2}} 
         \, e^{-x^2/C_i} 
               \; ,    \label{eq:DTplanes}
\end{equation}
where $x$ is the distance from the plane, $d_p$ the distance between neighboring planes, and $n$ the atomic density.

For a particle of mass $M$ and energy $E$ whose motion is governed by the continuum potential, the $z$-component of the force exerted by the crystal vanishes. Hence its longitudinal momentum $p_z$ will be a constant of motion. In consequence also the "longitudinal energy" $E_z \equiv (p_z^2c^2 + M^2c^4)^{1/2}$ as well as its "transverse" counterpart $E_{\perp} \equiv E-E_z$ are conserved. The latter is composed of the potential energy belonging to the interaction with the continuum crystal and kinetic energy associated with the transverse motion. $\it{Channeling}$ corresponds in our setup to bound motion of the electrons around a single string of atoms respectively of the positrons between a set of adjacent planes, that is, to the kinetic energy associated with the transverse motion being less than the depth or height $U_0$ of the continuum potential. For GeV electrons and positrons quantum states are close-lying and the motion is effectively classical. This is revealed by an estimate of the number of bound states of transverse motion. The number of bound states is proportional to the Lorentz factor $\gamma$ in the axial case and to $\gamma^{1/2}$ in the planar case, where the constant of proportionality depends on the target material and is of order 1 for electrons but larger for positrons% \cite{Sorensen1996}}}
. In the current investigation $\gamma\sim 10^5$.
For a thorough discussion of the motion of charged particles in aligned single crystals, including details on the differences and similarities between planar and axial channeling, the reader is refered to the original publication by J. Lindhard \cite{Lind65}, the extensive lecture notes by J. U. Andersen \cite{JUAnotes}, as well as review articles \cite{Sorensen1996, Ugge05}.

The ``critical angle'' or ``Lindhard angle'' provides a measure for incidence angles to crystal axes or planes below which a high fraction of the incoming particles will be channeled. In the axial case, where it is usually denoted $\psi_1$,  the critical angle assumes the value
\begin{equation}
 \psi_1\; = \; \sqrt{ \frac{4Z e^2}{pvd}}\; = \;  \frac{\alpha}{\sqrt{\gamma}\beta}\sqrt{\frac{4Za_0}{d}}
       \;  \label{eq:psi1}
\end{equation}
for unit-charge impact at momentum $p$ and velocity $v=\beta c$ on a target of atomic number $Z$. In the planar case the corresponding expression is
\begin{equation}
   \psi_p\; = \; \sqrt{ \frac{4Z e^2nd_pCa}{pv}}\; = \; 
             \frac{\alpha}{\sqrt{\gamma}\beta}\sqrt{4Za_0nd_pCa}
       \; , \label{eq:psip}
\end{equation}
where $C^2$ is a constant normally set to 3 and $a$ is the atomic screening length usually chosen as the Thomas-Fermi value $a=0.885a_0Z^{-1/3}$. Note that $\psi_1$ and $\psi_p$ both scale as $1/(pv)^{1/2}$, that is, for high values of the Lorentz factor they decrease in proportion to $1/\sqrt{\gamma}$. Critical angles for different energy and crystal combinations relevant in this investigation are listed in Table \ref{tb:Lindhard_angle}. 

\begin{table}[t]
	\begin{tabularx}{\linewidth}{p{0.25\linewidth} p{0.25\linewidth} p{0.25\linewidth} p{0.25\linewidth}}
		Crystal 								 & Energy & Critical angle 		   & $\Theta_B$ \\ \hline
		\multirow{2}{*}{C $\langle 100 \rangle$} & 40 GeV & $50~\mu\text{rad}$ & \multirow{2}{*}{$175~\mu$rad} \\
												 & 80 GeV & $35~\mu\text{rad}$ & \\ 
		Si (110)								 & 50 GeV & $23~\mu\text{rad}$ & $45~\mu$rad \\
	\end{tabularx}
	\caption{Critical Lindhard angle $\psi_p$ (plane), $\psi_1$ (axis) and Baier angle $\Theta_B$ for the energy and crystal combinations used in the experiment.}
	\label{tb:Lindhard_angle}
\end{table}

The motion of a penetrating charged particle according to the continuum model is perturbed by the difference between the true lattice potential, accounting for discreteness of target constituents and fluctuations in position due to thermal vibrations and quantum behavior, and the continuum potential. In the simulations below we account for scattering on individual target atoms and electrons by adding, at each time step in the integration, a velocity change selected at random according to a Gaussian distribution of width proportional to the root-mean-square multiple scattering angle pertaining to the distance traveled during the time step (the factor of proportionality chosen is $1/2$, see end of Subsec.\ \ref{subsec:rad}). As in other simulations, e.g. \cite{Koro13}, the deflection in the continuum potential is neglected in the calculation of this scattering. The error committed by such omission appears for diffraction and bremsstrahlung in oriented single crystals in the quantum perturbation limit as a reduction by 10--20\% below the amorphous yield of the incoherent contribution, %, due to ``thermal diffuse scattering'', 
see e.g. \cite{Sorensen1996} (the reduction factor, typically 0.8--0.9, is 1 minus a Debye-Waller factor in the Born approximation). The unsystematic or incoherent scattering is presumed to follow the local density of the scatterer since multiple scattering is dominated by close collisions %. For scattering on target electrons, the application of the local density may at first seem somewhat questionable but it remains a reasonable approximation since distant collisions between the projectile and target electrons only perturb the channeling or near-channeling motion marginally
 \cite{JUAnotes}. 

With the assumption of uncorrelated electron and nuclear contributions to multiple scattering, the mean-square scattering angle is the sum of the mean-square angles belonging to each of the two types of collisions. By introduction of the radiation length $X_0$ corresponding to the (average) density of target atoms, the result derived in \cite{Jackson_b_1975} for the nuclear contribution over a distance $\Delta l$ may, for projectiles of unit charge and velocities near $c$, be expressed as
\begin{equation}
\label{eq:MSnucl}
    \frac{\langle\theta^2\rangle_n}{\Delta l}=\frac{4\pi\alpha^{-1}m^2c^4}{E^2}\frac{1}{X_0}\frac{n_n({\bf{r}})}{n},
\end{equation}
where the constant in the numerator equals (21.2 MeV)$^2$. For a single transverse direction, the mean-square scattering angle is half the value displayed in Eq.\ (\ref{eq:MSnucl}) corresponding to a constant of (15.0 MeV)$^2$. When applied in Eq.\ (\ref{eq:MSnucl}) the radiation length should not include the electron contribution, that is,
\begin{equation}
\label{eq:X0}
    \frac{1}{X_0}=4\alpha r_e^2 n Z^2 \ln{(184Z^{-1/3})},
\end{equation}
where $Z$ is the target atomic number and the argument of the logarithm is according to \cite{PDG_2018}. 
It may be noted that the formula quoted in \cite{PDG_2018} for the root-mean-square multiple scattering angle in one dimension contains a slightly different constant than the 15.0 MeV corresponding to the expression (\ref{eq:MSnucl}), namely 13.6 MeV. As explained in \cite{LynchDahl1991}, fits to numerical simulations produce 13.6 MeV along with a depth dependence of the root-mean-square scattering angle containing a logarithmic term. Our simplified implementation of the multiple scattering is inconsistent with such dependence. We therefore stick to the expression (\ref{eq:MSnucl}), the error committed being at the same level as those due to the other approximations made in the simulations.

For impact of a heavy particle of unit charge at high $\gamma$, the electronic contribution to the mean-square multiple scattering angle (two independent transverse directions) attains the value \cite{Taratin1998}
\begin{equation}
\label{eq:MSel}
    \frac{\langle\theta^2\rangle_{el}}{\Delta l}=\frac{4\pi r_e^2}{\gamma^2}\left[ 
          \ln\left(\frac{2m\gamma^2c^2}{I}-1 \right) \right] n_{el}({\bf{r}}) ,
\end{equation}
where $I$ is the mean ionization potential.
In this form, the average of the squared scattering angle due to collisions with target electrons has been expressed in terms of the electronic stopping power. Restriction to close collisions is included by division by a factor of 2 reflecting the equipartition between close and distant collisions in electronic stopping (see also \cite{Lind65} and \cite{EBblue}).
While it is standard to apply (\ref{eq:MSel}) the expression may appear odd since it contains the local electron density along with the global electronic stopping logarithm (from Bethe's expression for the stopping power) which corresponds to a maximum range of interaction increasing with $\gamma$, reaching far beyond a single channel at the energies considered in our study. Furthermore, at these energies the stopping logarithm is actually smaller than that appearing in (\ref{eq:MSel}) due to polarization of the target electron gas, the so-called density effect. But without the density effect there is equipartition between close and distant collisions for heavy particles and hence Eq. (\ref{eq:MSel}) applies. For electron or positron impact the electronic contribution will actually be less % than given by Eq. (\ref{eq:MSel})
since, due to kinematics, close collisions contribute less than %half to the stopping power without the density effect. 
for heavy particles, in our case approximately 1/3 less. We have ignored this difference and applied Eq. (\ref{eq:MSel}) for the electronic contribution to the multiple scattering. The electronic contribution is generally of minor importance except for ``proper'' channeled positrons which move through the crystal in open channels far from target nuclei. Such proper channeled positrons constitute a small fraction of all particles, even for the most restrictive cuts.

In the harmonic approximation for interatomic forces the local density of target nuclei to enter Eq.\ (\ref{eq:MSnucl}) reads 
\begin{equation}
\label{eq:nnuclstring}
    n_n\!\!\!^{\text{string}}(r)=\frac{1}{\pi\rho^2d}e^{-r^2/\rho^2}
\end{equation}
for an isolated row of atoms and
\begin{equation}
\label{eq:nnuclplane}
    n_n\!\!\!^{\text{plane}}(x)=\frac{nd_p}{\sqrt{\pi}\rho}e^{-x^2/\rho^2}
\end{equation}
for a single plane. The same distributions were applied in the derivation of the expressions (\ref{eq:DT}--\ref{eq:DTplanes}) for the continuum potential.
The local electron density may be obtained from Doyle and Turner's fit to the X-ray scattering factor \cite{Doyl68}. For an isolated string of atoms the density is given as \cite{Jens2013}
\begin{equation}
\label{eq:nelstring}
   n_{el}\!\!\!^{\text{string}}(r)  = 
         \frac{1}{\pi d}\sum_{i=1}^4\frac{a^{(X)}_i}{C^{(X)}_i} e^{-{r^2/C^{(X)}_i}}  
          +c^{(X)} n_n\!\!\!^{\text{string}}(r) ,    
\end{equation}
where 
\begin{equation}
\label{eq:DTCXi}
  C^{(X)}_i=C^{(X)}_i(\rho )\equiv b^{(X)}_i/4\pi^2+\rho^2
            \; .     
\end{equation}
For an isolated plane, the expression is
\begin{equation}
\label{eq:nelstring}
   n_{el}\!\!\!^{\text{plane}}(x)  = 
         \frac{nd_p}{\sqrt{\pi}}\sum_{i=1}^4\frac{a^{(X)}_i}{\sqrt{C^{(X)}_i}} e^{-{x^2/C^{(X)}_i}}  
          +c^{(X)} n_n\!\!\!^{\text{plane}}(x) .    
\end{equation}
For details and explicit values of the coefficients $c^{(X)}$, $a^{(X)}_i$, and $b^{(X)}_i$ see \cite{Doyl68}. Note that $c^{(X)}+\sum_{i=1}^4a^{(X)}_i=Z$. 

\subsection{Radiation}
\label{subsec:rad}
In our computational analysis each individual positron or electron follows a classical trajectory characterized by the instantaneous position $\bm{r}(t)$, the instantaneous velocity $\bm{v}(t)=\bm{\beta}(t)c$, and the instantaneous acceleration $\dot{\bm{v}}(t)=\dot{\bm{\beta}}(t)c$. According to classical electrodynamics, the electromagnetic energy $E_{\gamma\text{c}}$ radiated  per unit of frequency $\omega$ and of solid angle $\Omega$ is given by \cite{Jackson_b_1975}
\begin{equation}
\small
\label{dE_dodO}
\frac{d^2E_{\gamma\text{c}}}{d\omega d\Omega}=\frac{e^2}{4\pi^2c}\left|\int_{-\infty}^{\infty} \frac{\bm{n}\times[(\bm{n}-\bm{\beta})\times\dot{\bm{\beta}}]}{(1-\bm{n}\cdot\bm{\beta})^2}e^{i\omega(t-\bm{n}\cdot\bm{r}/c)}dt\right|^2 ,
\end{equation}
where $\bm{n}=(\sin\vartheta\cos\varphi,\sin\vartheta\sin\varphi,\cos\vartheta)$ is the direction of emission with polar and azimuthal angles $\vartheta$ and $\varphi$ defined relative to a suitable axis and $d\Omega=\sin\vartheta d\vartheta d\varphi$. Equation (\ref{dE_dodO}) is in accordance with the relativistic generalization of the Larmor formula (the Li\'{e}nard formula) for the radiated power \cite{Jackson_b_1975}.  

If the only non-negligible quantum effect is the photon recoil it can be taken into account by a simple substitution of the frequency variable in the classical photon number spectrum regardless of the details of the motion of the particle \cite{Lind91}. Based on the Weizs{\"a}cker-Williams method \cite{Jackson_b_1975,WeizsackerWilliams1934}, but bypassing the actual computation of the virtual photon spectrum, Lindhard showed that for spin-0 particles with energy $E$, substituting the frequency $\omega$ by
\begin{equation}
 \omega^* = \omega/(1-\hbar\omega/E)
\label{eq:omegastar}
\end{equation}
in the classical number spectrum gives exactly the quantum number spectrum for single-photon emission \cite{Lind91}
\begin{equation}
\frac{dN_\text{c}}{d\hbar\omega}(\omega^*) = \frac{dN_\text{q}}{d\hbar\omega}(\omega) \; .
\end{equation}
This implies that the intensity spectrum translates as
\begin{equation}
\label{Lindhard_substi}
\frac{dE_{\gamma\text{q}}}{d\hbar\omega}(\omega) = \frac{\omega}{\omega^*}\frac{dE_{\gamma\text{c}}}{d\hbar\omega}(\omega^*).
\end{equation}
The result (\ref{Lindhard_substi}) is confirmed by a full quantum mechanical calculation for a spinless particle moving in a constant magnetic field carried out many years prior to Lindhard's investigation \cite{Matveev1957}. Therefore, when effects related to the spin of the positrons or the electrons can be neglected, employing the substitution of the frequency variable in Eq.\ (\ref{dE_dodO}) according to the prescription Eq.\ (\ref{Lindhard_substi}) will reproduce the full quantum spectrum. We call this model the substitution model.

To include the effects of both the quantum recoil and the spin in the radiation process, we apply a result obtained based on the semi-classical method by Baier and Katkov \cite{Baie98}, in which the particle motion is treated classically whereas the interaction with the radiation field is quantal and to first order. For a general trajectory Belkacem, Cue and Kimball, give the spectrum as  \cite{belkacem_1985}: 
\begin{equation}
\label{cue}
\frac{d^2E_\gamma}{d\hbar\omega d\Omega}=\frac{\alpha}{4\pi^2}\left(\frac{E^{*2}+E^2}{2E^2}\left| I \right|^2 + \frac{(\hbar\omega)^2}{2E^2\gamma^2}\left| J \right|^2\right),
\end{equation}
where $E^* = E-\hbar\omega$, ($E^*\omega^*=E\omega$) and $I$ and $J$ are given by
\begin{equation}
\label{Iterm}
I = \int_{-\infty}^{\infty} \frac{\bm{n}\times[(\bm{n}-\bm{\beta})\times\dot{\bm{\beta}}]}{(1-\bm{n}\cdot\bm{\beta})^2}e^{i\omega^*(t-\bm{n}\cdot\bm{r}/c)}dt,
\end{equation}
\begin{equation}
\label{Jterm}
J = \int_{-\infty}^{\infty} \frac{\bm{n}\cdot\dot{\bm{\beta}}}{(1-\bm{n}\cdot\bm{\beta})^2}e^{i\omega^*(t-\bm{n}\cdot\bm{r}/c)}dt .
\end{equation}
For derivations, see \cite{kimball_1986} and \cite{Tobias_2015}.
We call this the BCK model.

The radiation resulting from the unsystematic scattering on target constituents, which we account for in our model through the inclusion of multiple scattering in the determination of the trajectory entering the radiation integrals Eqs. (\ref{dE_dodO}), (\ref{Iterm}--\ref{Jterm}), poses a separate challenge. For a small change in velocity happening over a distance sufficiently short that the variation of the phase in the exponential factor in the intensity (\ref{dE_dodO}) is negligible, that is, for a distance short compared to the formation length, the radiated intensity amounts to
\begin{equation}
\label{eq:deltav2}
\frac{dE_\gamma}{d\omega} = \frac{2}{3\pi}\frac{e^2}{c}\gamma^2\vert {\bm{\Delta\beta}}\vert^2
\end{equation}
according to classical electrodynamics \cite{Jackson_b_1975}. The result holds for $\vert {\bm{\Delta\beta}}\vert$ smaller than, approximately, $2/\gamma$ (dipole regime) beyond which the intensity becomes small. A derivation based on the BCK formula (\ref{cue}) yields \cite{Tobias_2015}
\begin{equation}
\label{eq:deltav2rel}
\frac{dE_\gamma}{d\hbar\omega} = \frac{2}{3\pi}\alpha\gamma^2\vert {\bm{\Delta\beta}}\vert^2
      \left( 1-\frac{\hbar\omega}{E}+\frac{3}{4}\left( \frac{\hbar\omega}{E}\right)^2\right) ,
\end{equation}
which reduces to (\ref{eq:deltav2}) in the classical limit.
In the simulations, the distance travelled during one time step is much smaller than the formation length. The time steps used are such that essentially all velocity changes computed according to the multiple scattering distributions in one time step 
are less than $2/\gamma$ and hence contribute to the radiation according to Eq.\ (\ref{eq:deltav2}) or (\ref{eq:deltav2rel}). However, in the collision with a single atom, the mean-square scattering angle is about twice the average of the square of the scattering angles contributing to bremsstrahlung. 
In other words, if $\vert {\bm{\Delta\beta}}\vert^2$ in (\ref{eq:deltav2rel}) is replaced by $\langle\theta^2\rangle$ as given by Eq. (\ref{eq:MSnucl}), a radiation intensity twice that given by the Bethe-Heitler formula for bremsstrahlung results.
Hence the radiation associated with unsystematic scattering is overestimated by a factor of 2, essentially, in our implementation. 
See also \cite{Bell1958} for a similar note. To account for this, we reduce the multiple scattering, Eqs.\ (\ref{eq:MSnucl}) and (\ref{eq:MSel}), by a factor of 2. This reduction may be seen as a way to compensate for the dipole approximation implicitly imposed by our numerical procedure to the multiple-scattering contribution to the radiation. Our simulations then produce the correct high-energy Bethe-Heitler tails of the radiation spectra visible beyond the coherent contribution pertaining to motion in the continuum potential. The reduced redistribution over "transverse energies" is expected to have minor influence on the radiation spectra.

\subsection{Quantum reduction of the damping force}
\label{subsec:CFA}
Generally, the energy radiated by an electron or a positron following a classical trajectory is higher according to classical than to quantum electrodynamics. Hence, the LL equation overestimates the radiation reaction in case quantum effects are of any influence. A priori there is no cure for this since a quantum version of the LL equation does not exist. However, under certain circumstances, the emission process appears as if the electromagnetic field were constant, and for such a field the ratio of the radiated energy, quantum to classical, is known. This may be utilized to reduce the damping force contained in the LL equation, albeit in an approximate manner.

When the variation in angle of a charged particle passing through the crystal is large compared to the opening angle of the radiation cone, $1/\gamma$, the emitted radiation approaches that pertaining to a (locally) constant electromagnetic field. The critical channeling angle decreases slower than the opening angle of the radiation cone with increasing energy. At high energies, a charge traversing the crystal far beyond the channeling region will experience deflections beyond $1/\gamma$ when incident with angles to the plane/axis less than $\Theta_B= U_0/m$ \cite{Sorensen1996}. For a silicon crystal oriented along the (110) plane $\Theta_B$ is $\simeq$ 45 $\mu$rad, while for a diamond crystal oriented along the $\langle 100 \rangle$ axis of diamond we find  $\Theta_B\simeq$ 175 $\mu$rad. In the positron experiment, $\Theta_B$ is only twice the critical angle and only four times the opening angle of the radiation cone, so a constant field approximation (CFA) is dubious here as also discussed in \cite{cern2017}. In the electron experiment, on the other hand, $\Theta_B$ is around 3.5 (5) times the critical angle at 40 GeV (80 GeV) and 14 (28) times the opening angle of the radiation cone, so the CFA is expected to work quite well in and near the axial channeling regime. 

As a measure of the applicability of the CFA under channeling conditions, the authors of \cite{Baie98} introduced the parameter $\rho_c=2U_0E/m^2$ (termed $\xi^2$  in the strong-field laser community and known as the dipolarity parameter). %With $\rho_c$ being defined as twice the square of the ratio of deflection angle to $1/\gamma$  it 
It reduces to $ \rho_c = 2\Theta_B/\gamma^{-1}$ which assumes a value close to 9 for planar channeling of 50 GeV positrons and around 55 (27) for 80 GeV (40 GeV) electrons axially channeled. The large value of $\rho_c$ in the latter case again indicates that CFA is a very useful approach for channeled electrons in our experiment, in particular at the highest energy, whereas its modest value in the former case points towards a limited use of CFA for the positrons as also shown in \cite{cern2017} by explicit calculation of spectra.

A quantum correction to the intensity of the emitted radiation inherent in the LL equation may be included in an approximate manner for the axial case by multiplying the radiation reaction force contained in the LL equation by a damping factor defined as  the ratio between the quantum and classical radiation intensity in the constant field approximation. An approximate expression for this factor is
\begin{equation}\label{eq:gchi}
G(\chi) = \left[1 + 4.8(1+\chi)\ln(1+1.7\chi)+2.44\chi^2\right]^{-2/3} \; .
\end{equation}
The accuracy of the analytical expression (\ref{eq:gchi}) is within 2\% for all values of $\chi$ \cite{Baie98}, and $G(\chi)$ has been validated experimentally in \cite{kandersen2012}. 
\begin{figure*}[t]
	\centering
	\begin{subfigure}[t]{0.5\textwidth}
		\centering
		\includegraphics[width=1.0\columnwidth]{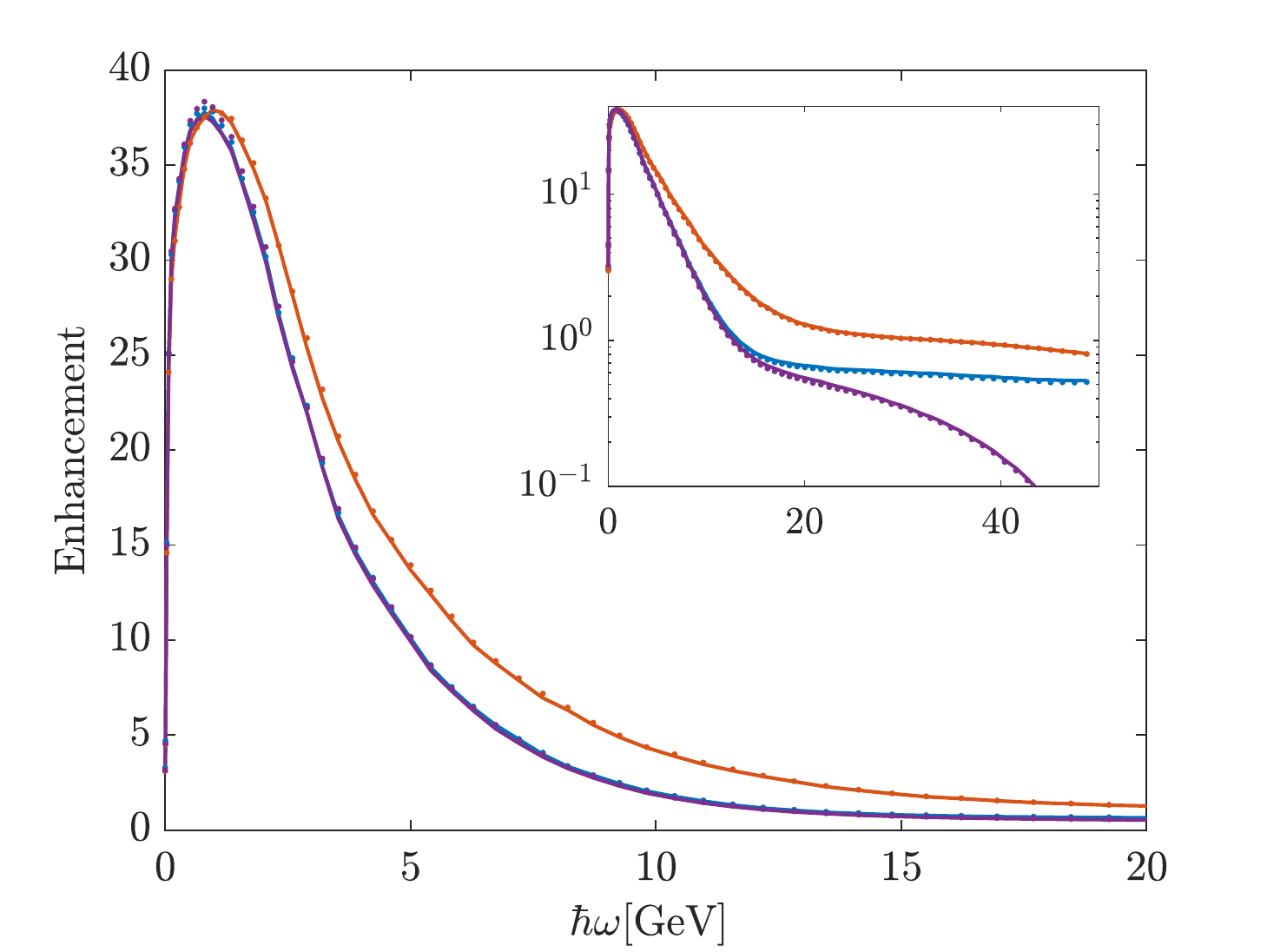}
		\caption{}
	\end{subfigure}%
	~ 
	\begin{subfigure}[t]{0.5\textwidth}
		\centering
		\includegraphics[width=1.0\columnwidth]{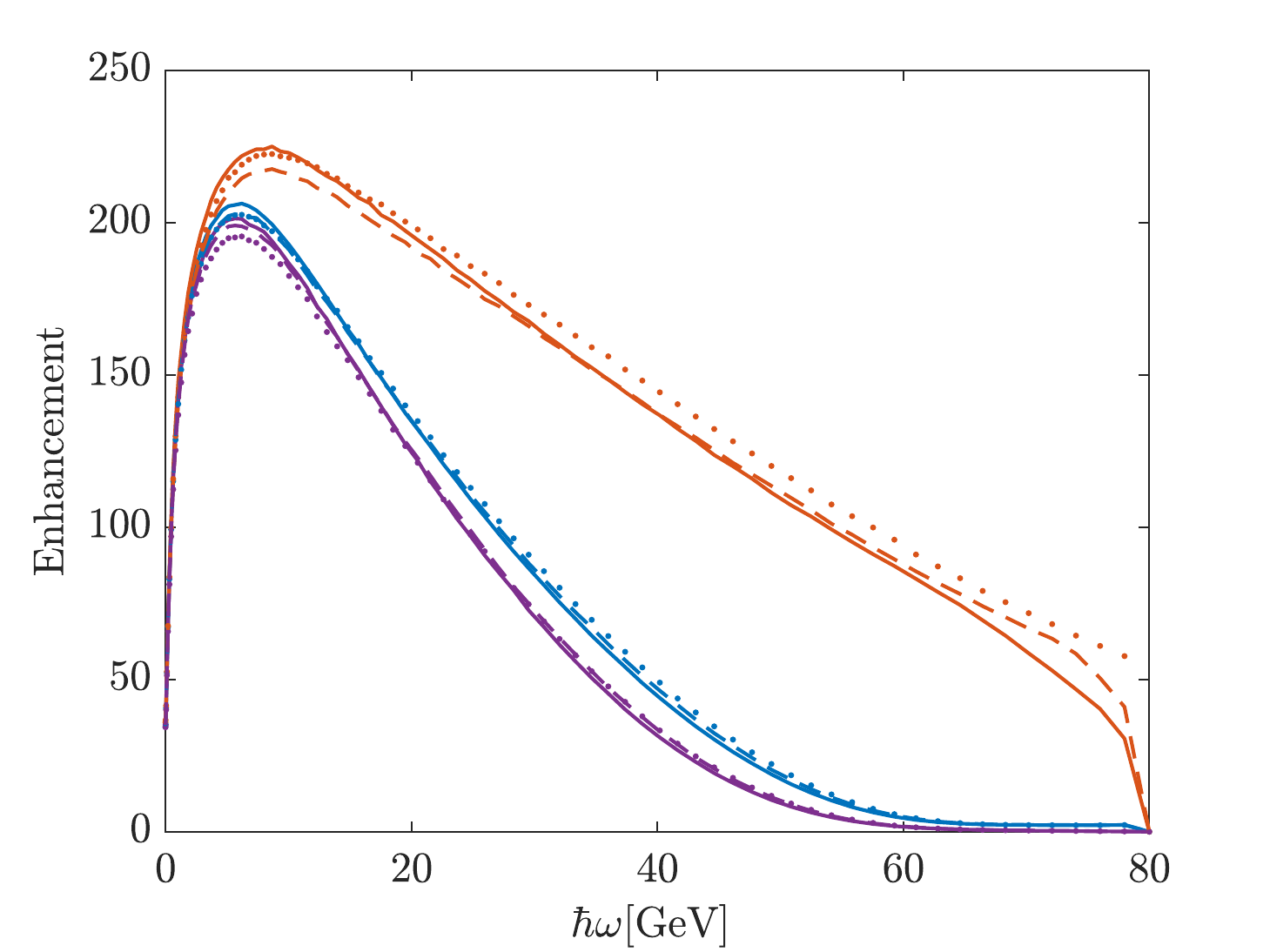}
		\caption{}
	\end{subfigure}
		\caption{(a) Theoretical enhancement spectra for 50 GeV positrons channeled in the (110) plane of a 0.1mm thick Si crystal. Full-drawn spectra pertain to trajectories determined by the LL equation of motion. Dotted spectra pertain to trajectories determined by the Lorentz force. Spectra obtained in the classical model are drawn in red, spectra obtained with the substitution model appear in purple, BCK spectra are drawn in blue. The insert shows the spectra on a logarithmic intensity scale all way the up to the primary energy. (b) Theoretical enhancement spectra for 80 GeV electrons channeled along the $\langle 100 \rangle$ axis of a 20$\mu$m thick diamond single crystal. Spectra obtained using the classical model are drawn in red, spectra obtained using the BCK model in the CES are drawn in blue, and spectra obtained using the substitution model in the CES are drawn in purple. Dotted lines show spectra obtained using pure Lorentz-force trajectories, solid lines indicate the LL equation has been used, and dashed lines are spectra calculated for trajectories determined by the LL equation including the $G(\chi)$-correction.} 
	\label{fig:ThinCrystals}
\end{figure*}

\subsection{Implementation}
\label{subsec:implementation}
\begin{figure*}[t]
	\centering
	\begin{subfigure}[t]{0.5\textwidth}
		\centering
		\includegraphics[width=1.0\columnwidth]{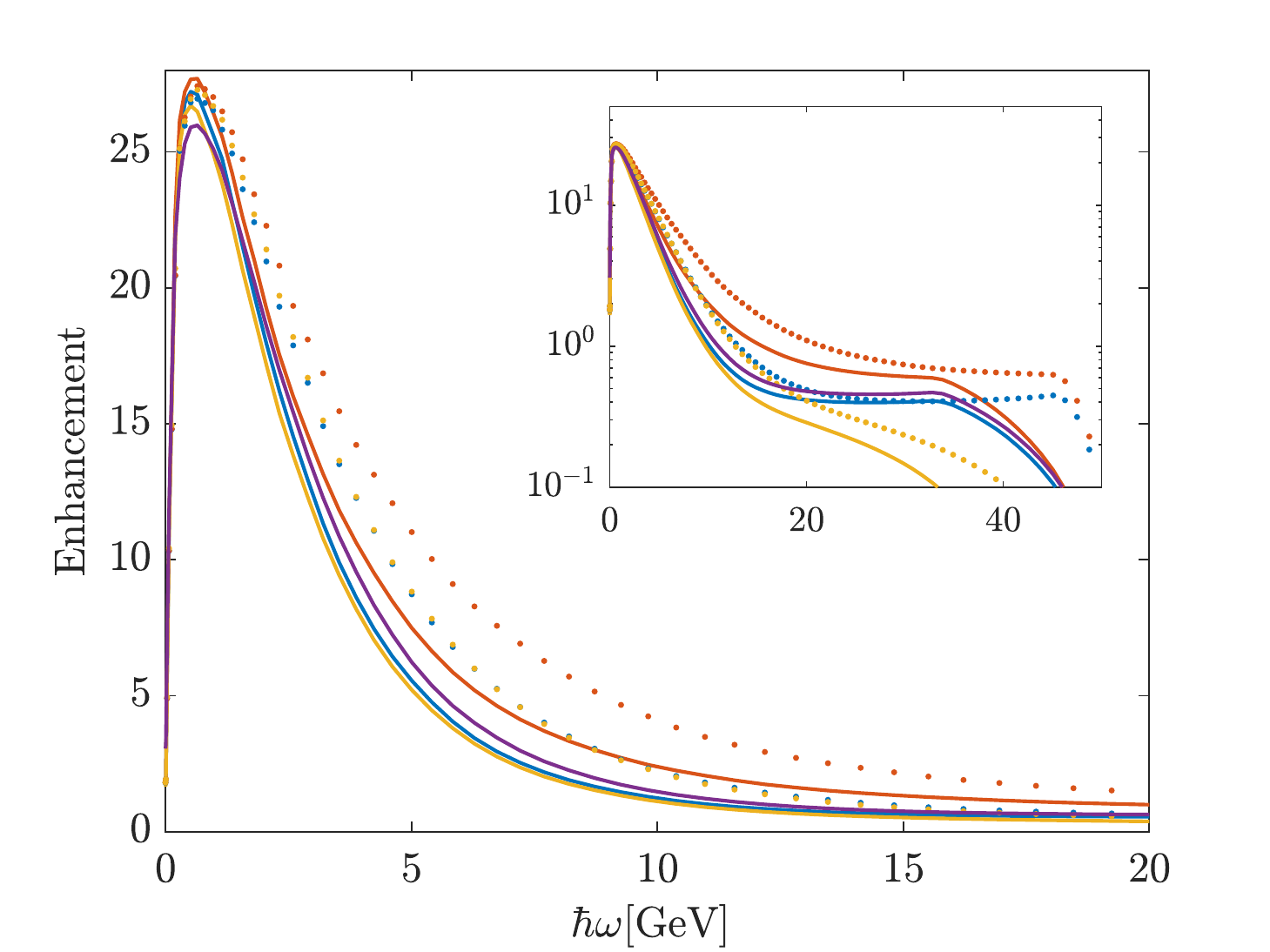}
		\caption{}
	\end{subfigure}%
	~ 
	\begin{subfigure}[t]{0.5\textwidth}
		\includegraphics[width=1.0\columnwidth]{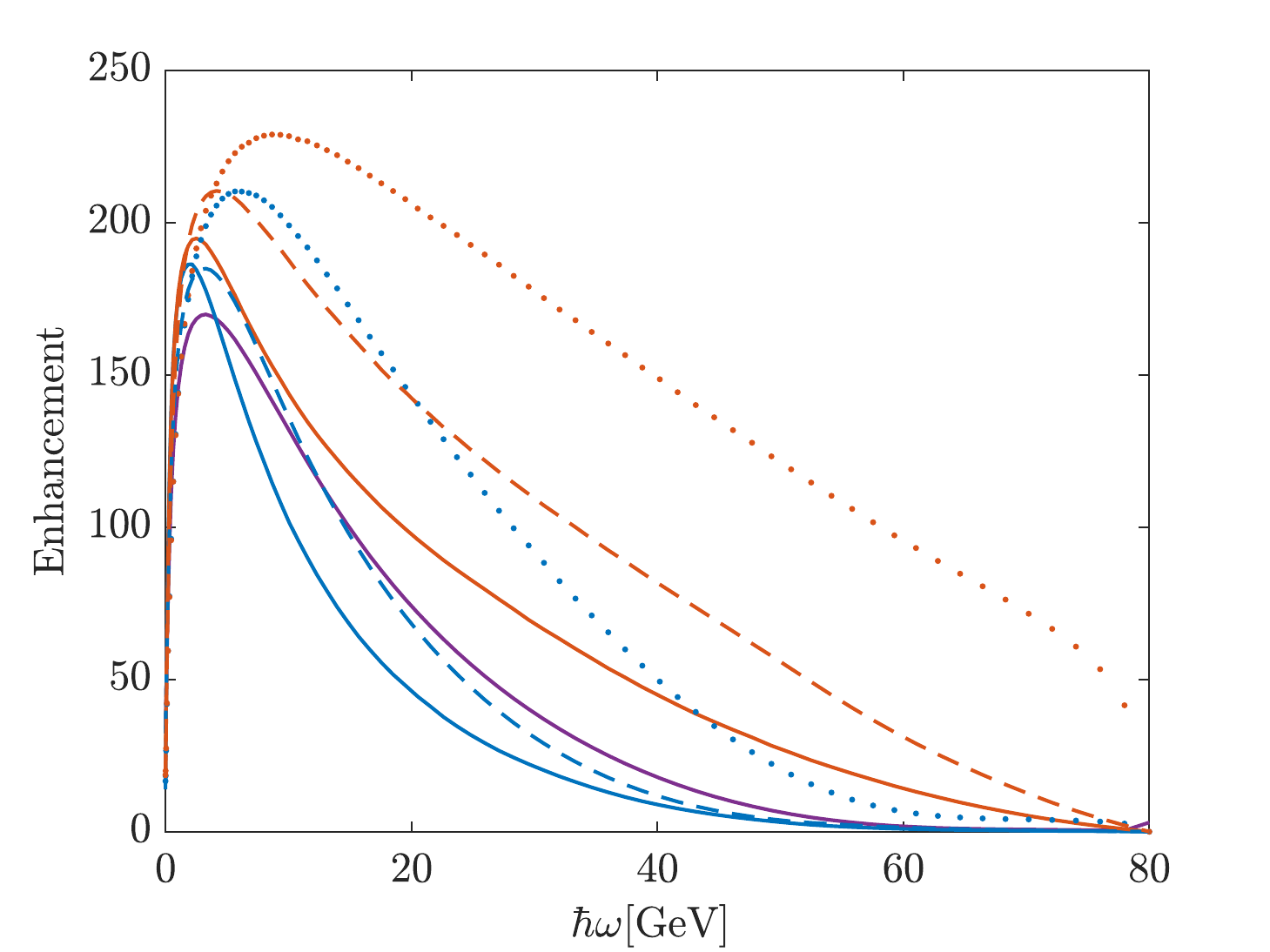}
		\caption{}
	\end{subfigure}
	\caption{(a) Theoretical enhancement spectra for 50 GeV positrons channeled in the (110) plane of a 6.2mm thick Si single crystal. Spectra drawn in solid lines, except purple, pertain to trajectories determined by the LL equation of motion. Dotted spectra pertain to trajectories determined by the Lorentz force. Spectra obtained in the classical model are drawn in red, spectra obtained with the substitution model using the CES appear in yellow, BCK spectra using the CES are drawn in blue. The purple spectrum is calculated by the application of the stochastic scheme. The insert shows the spectra on a logarithmic intensity scale all the way up to the primary energy.
The quantum models (yellow, blue and purple) overlap in the coherent region where they are hard to distinguish. (b) Theoretical enhancement spectra for 80 GeV electrons channeled along the $\langle100\rangle$ crystallographic axis of a 1.0mm thick diamond single crystal. Spectra obtained in the classical model are drawn in red, spectra from the BCK model using the CES appear in blue, and a spectrum simulated in the stochastic scheme is drawn in purple. Pure Lorentz-force trajectories have been used for the dotted spectra, the LL equation has been used for solid-line spectra, except purple, and dashed lines display spectra based on the LL equation with the $G(\chi)$-correction.}
	\label{fig:models_compare}
\end{figure*}
\begin{table*}[t]
	\begin{tabularx}{\linewidth}{p{0.082\textwidth} p{0.082\textwidth} p{0.082\textwidth} p{0.14\textwidth} p{0.082\textwidth} p{0.082\textwidth} p{0.082\textwidth} p{0.082\textwidth} p{0.082\textwidth} p{0.082\textwidth}}
		Crystal & $d_c$ & E & Cut & $\overline{\chi}$ & $\sqrt{\overline{\chi^2}}$ & $\Delta$& $\Delta E_{\mathrm{LL}}$ & $r_\mathrm{LL}$ & $r_{\mathrm{LL, G(\chi)}}$ \\ \hline
		% C
		\multirow{12}{*}{C $\langle 100 \rangle$} & \multirow{6}{*}{1.0 mm} &  \multirow{3}{*}{40 GeV } & No cut & 0.0305 & 0.050 & 40\% & 26\%& 0.52 & 0.8 \\
		& & & $2\psi_1 < \psi < 4\psi_1$& 0.0293& 0.048 & 37\% & 24\%& 0.5 & 0.76 \\
		& & & $\psi_1 > \psi$ & 0.0337&  0.054&   47\%& 20\%&  0.59 &  0.91 \\
		& &  \multirow{3}{*}{80 GeV } & No cut & 0.0563& 0.091 &  68\% & 34\%&  0.40 & 0.75 \\
		& & & $2\psi_1 < \psi < 4\psi_1$& 0.0551&  0.090&  66\%& 33\%& 0.58 & 0.78\\
		& & &  $\psi_1 > \psi$ & 0.0640&  0.101  &   83\%& 42\%& 0.49 & 0.93 \\
		&  \multirow{6}{*}{1.5 mm} &  \multirow{3}{*}{40 GeV } & No cut & 0.0282& 0.046  &   52\%& 34\%& 0.54 & 0.80 \\
		& & & $2\psi_1 < \psi < 4\psi_1$& 0.0277& 0.046 &  50\%&33\%& 0.52 & 0.77 \\
		& & &  $\psi_1 > \psi$ & 0.0311&  0.050&     61\%& 41\%& 0.60 & 0.91 \\
		& &  \multirow{3}{*}{80 GeV } & No cut & 0.0511&0.084&    86\%& 45\%& 0.42 & 0.75 \\
		& & & $2\psi_1 < \psi < 4\psi_1$& 0.0505&  0.084&     85\%& 44\%& 0.43 & 0.77 \\
		& & & $\psi_1 > \psi$ & 0.0576& 0.093&    104\%& 54\%& 0.51& 0.93 \\ \hline
		% Si
		\multirow{8}{*}{Si (110)} & \multirow{2}{*}{1.1 mm} & \multirow{8}{*}{50 GeV} & No cut & 0.0155&  0.021&   5\%& 6\%&  0.66 & 0.74 \\
		&  &  & $\psi < 30\mu$rad & 0.0140&  0.017&    4\%& 5\%& 0.84 & 0.94 \\
		& \multirow{2}{*}{2.0 mm} & & No cut & 0.0155&  0.020&     9\%& 12\%& 0.67 & 0.75 \\
		&  &  & $\psi < 30\mu$rad & 0.0130&  0.017&     7\%& 9\%& 0.84 & 0.94 \\
		& \multirow{2}{*}{4.2 mm} & & No cut & 0.0143& 0.017&     17\% & 21\% & 0.68 & 0.75 \\
		&  &  & $\psi < 30\mu$rad & 0.0124&  0.014&    13\%& 16\%& 0.83 & 0.93\\
		& \multirow{2}{*}{6.2 mm} & & No cut & 0.0139& 0.017&    24\%& 30\% & 0.71 & 0.78 \\
		&  &  & $\psi < 30\mu$rad & 0.0116&  0.014&     17\%& 20\%& 0.84 & 0.93 \\
	\end{tabularx}
	\caption{The average strong-field parameter $\bar{\chi}$, the root mean square of the strong field parameter $\sqrt{\overline{\chi^2}}$, the expected energy loss $\Delta$ from \cref{eq:LLloss}, the average energy loss of the particles according to the trajectory $\Delta E_{\mathrm{LL}}$ for trajectories calculated using the LL force, the fractional difference between average energy lost according to the trajectory and according to the full spectrum  $r =\Delta E_{\mathrm{spectrum}} / \Delta E_\mathrm{trajectory}$ for trajectories calculated using the Lorentz equation including the LL force without and with the $G(\chi)$ correction, $r_{\mathrm{LL}}$ and $r_{\mathrm{LL, G(\chi)}}$ respectively. The spectral energy loss is calculated with the BCK model using the CES. Data is shown for the different cuts, crystals, crystal thicknesses $d_c$, and energies $E$, used in the experiment. The values of $\bar{\chi}$, $\sqrt{\overline{\chi^2}}$, $\Delta$, and $\Delta E_{\mathrm{LL}}$ are evaluated using the $G(\chi)$ correction for the electrons and the pure LL for the positrons, while $r_{\mathrm{LL, G(\chi)}}$ is calculated using the $G(\chi)$ correction and $r_{\mathrm{LL}}$ is calculated without for both positrons and electrons.}
	\label{tb:table2}
\end{table*}

%Tekst Implementation starter her:
The classical radiation formula, Eq.\ (\ref{dE_dodO}), may be applied for any trajectory including trajectories determined from the LL equation. This is not the case for the semi-classical models since they are first-order approximations in the interaction with the radiation field, that is, they only apply for single-photon emission. Hence the semiclassical models only allow for pure Lorentz-force trajectories. The electrons and positrons in our experiment emit multiple photons when traversing a thick crystal. To adapt the semi-classical models to this scenario, we divide the crystal in sufficiently thin sections so that the probability for photon emission in each section is low. The radiation spectrum is evaluated in each section and the exit position and momentum of a particle in one section determine its initial conditions for the next section. Two schemes are applied. One, which comes in a number of variants, is based on the LL equation. The other, which serves as a benchmark for the first, is not based on the LL equation, but accounts for quantum stochasticity. The stochastic scheme is almost identical to that presented in \cite{cern2017}. 
A minimum photon energy of 1 GeV is taken in this scheme.

When dividing the crystal into smaller sections it is important that the section length is longer than the formation length $l_f =2\gamma^2(E-\omega)/E\omega$ of most photons emitted by the particle. For a positron or an electron of 40 to 80 GeV emitting a 5 GeV photon we have $l_f$ of around 0.4 to 2 $\mu$m. A section length of 20 $\mu$m is used for the axial case, which is significantly longer than the formation length of a typical photon in our experiment, but short enough that the emission probability is small. For 40 GeV electrons, only photons of energy less than 120 MeV have formation lengths longer than 20 $\mu$m, while for 80 GeV the limit is 0.5 GeV. In the planar case a similar analysis has led to the choice of a section length of 0.1 mm, which corresponds to the formation length of 40 MeV photons.

The radiative losses during the penetration of a section, albeit small, imply a time dependence of the energy of the radiating particle and, hence, of the frequency $\omega^*$, Eq.\ (\ref{eq:omegastar}). Since $\omega^*$ appears in the exponential phase factor of the radiation integrals, the spectrum then depends on the initial phase. This is nonphysical. We eliminate the  ambiguity by fixing the energy of the particle, where explicit in the radiation integrals, to its initial value when entering the section. We refer to this procedure as the Constant Energy Scheme (CES). Despite the name, the particle trajectory in the section is determined by the LL equation. 

The gradual radiative energy loss due to coherent scattering on crystal atoms, that is, due to the deflection in the continuum potential, is contained in the LL equation. Energy loss due to radiation emission associated with unsystematic scattering on individual nuclei and electrons is evaluated numerically in each integration step and subtracted. From Eq.\ (\ref{eq:deltav2rel}) the loss corresponding to a scattering angle $\theta$ is
\begin{equation}
\label{eq:ElossMSc}
\Delta E = \frac{1}{2\pi}\alpha\gamma^3mc^2\theta^2
\end{equation}
by integration over photon energies.

In the stochastic scheme a charged particle traverses a section governed only by the Lorentz force. The radiation spectrum from the charge is evaluated in each section using the BCK model and the probability for photon emission is calculated based on the spectrum. This probability and the spectrum is used to determine if a photon is emitted in the section and, if so, to fix its energy. If a photon is emitted, the particle loses an amount of energy equal to the energy of the emitted photon and continues into the next section. The corresponding change in particle momentum is obtained through the simplifying assumption that the photon is emitted in the direction of motion of the radiating particle, that is, only the magnitude of the momentum is changed, not the direction. The stochastic  scheme does not involve the LL equation. Compared to the implementation of the model in \cite{cern2017}, our implementation does not require the motion to be periodic, which is essential when considering electrons traversing axially aligned crystals.
%{\ahs{For other stochastic schemes see, e.g., \cite{Artru1990}, discussed in \cite{cern2017}, and \cite{Bandiera2013}--\cite{Bandiera2018} and references therein.}}
The stochastic scheme presented in \cite{Artru1990} is discussed in \cite{cern2017}.

As described in \cite{CFN2019GPU}, the task of evaluating the theoretical radiation spectra for particles penetrating oriented single crystals is not easy and, to our knowledge, not many programs exist which are capable of doing this. One first has to solve the equation of motion for particles moving in the continuum potential inside the crystal, and afterwards integrate each trajectory using one of the radiation integrals mentioned earlier. Since the radiation integrals are differential in energy ($d\hbar\omega$) and emission angle ($d\Omega$), we have to compute the integral for each emission angle in space for each photon energy to produce a full radiation spectrum. Each trajectory is influenced by stochastic perturbations due to multiple Coulomb scattering  and the initial conditions vary according to angle of incidence within the beam profile and point of entry into to crystal. As a result we have to make an average spectrum from thousands of individual incident particles before we can compare the simulated spectrum to an experimental spectrum.
A spectrum from a single beam of particles hitting a specific crystal is not unlikely to consist of $5\times 10^8$ integrals over the trajectory, which contains approximately $1\times 10^6$ points in time per mm crystal traversed.

In Sec.\ \ref{sec:results} the theoretical spectra are compared to the experimental spectra using two types of simulations. The first type of simulation produces a radiation spectrum from a clean beam penetrating an aligned crystal based on the theory described previously. Such theoretical spectra cannot be compared directly to experimental data because of the experimental environment smearing out the true spectrum emitted from within the crystal. Because of this we developed a second simulation which produces  a spectrum of the response of the experimental setup from the theoretical spectrum produced by the first type of simulation.  In this simulation, the response of the experimental setup is based on the geometry of the experiment, physics processes (multiple Coulomb scattering through the beamline elements, multiple Coulomb scattering of the pair created in the converter foil), the Mimosa-26 detector resolution, the conversion of two hits within $50~\mu$m into a single hit, and the opening ``Borsellino'' angle of the pair (of the order $mc^2/\hbar\omega$, with a known distribution). The output of the simulation is a data file identical to what we get from the experiment, this data file is then analyzed as with the experimental data files and a simulated spectrum is produced. The simulation of the experiment is also what produces the amorphous spectra we compare to the amorphous experimental spectra, here we use the Bethe-Heitler cross section when particles penetrate the crystal.

In some of our figures we display what we term an enhancement spectrum, which is the radiation spectrum emitted by particles traversing an aligned single crystal, divided by the radiation spectrum from particles with identical initial conditions traversing the same crystal placed in an amorphous orientation, that is, far from any major crystallographic axis or plane. For experimental radiation spectra, the enhancement spectra use the spectrum measured by orienting the crystal in an amorphous orientation for normalization. When simulated spectra are compared to experimental data, the simulated spectra are normalized to a bremsstrahlung spectrum simulated by substituting the cross-section for photon emission in the target crystal, in the simulation of the experiment, by the Bethe-Heitler cross-section \cite{PDG_2018}. Purely theoretical enhancement spectra are simply normalized to the analytical Bethe-Heitler radiation spectrum.

In Fig.\ \ref{fig:ThinCrystals}a we show enhancement spectra for 50 GeV positrons traversing a 0.1 mm thin silicon single crystal channeled in the (110) plane calculated by the CES. 
Particle entry angles are confined between $\pm 30$ $\mu$rad with respect to the plane, and sampled from a Gaussian distribution with a mean of 0 and divergence of 100 $\mu$rad. Since the radiative energy loss due to coherent scattering in this case amounts to a few GeV/mm, we see no effect of the LL equation. In all models, the spectrum produced in 0.1 mm of (110) silicon by trajectories governed by the LL equation is essentially indistinguishable from that where the positron motion is governed solely by the Lorentz force.

Comparing the different models, three important observations are apparent:
\begin{enumerate}
  \item Quantum corrections are observed near the peak of the enhancement spectrum which corresponds to coherent scattering on crystal constituents.  
  \item The substitution and BCK models produce essentially identical results up to 20 GeV, as expected.
  \item At high energy, where the coherent effects have vanished, the substitution model fails due to neglect of effects of the positron spin. 
\end{enumerate}
It should be noted that the level of the high-energy tail, which is due to incoherent scattering on target constituents, is below that pertaining to a uniform particle flux through the target due to focusing in low-density regions resulting from the channeling motion \cite{Lind65}. For the high-energy tail of the spectrum, the discrepancy between the classical and BCK model derives from the last factor in Eq.\ (\ref{eq:deltav2rel}), a signature of the Bethe-Heitler bremsstrahlung formula, which reduces to 3/4 in the high-energy limit.

In Fig.\ \ref{fig:ThinCrystals}b we show theoretical enhancement spectra for 80 GeV electrons traversing a 20 $\mu$m thin diamond single crystal with incident angles $\psi < \psi_1$ relative to the $\langle 100 \rangle$ axis. Under these conditions the radiative energy loss is $\sim 50$ GeV per mm. Again, we still see essentially no effect of the LL equation on the spectrum due to the small thickness. The discrepancy between the two semi-classical models, although moderate, points out that the spin-contribution cannot be neglected in this case. Due to the larger value of $\overline{\chi}$, 0.06 in Fig.\ \ref{fig:ThinCrystals}b compared to 0.015 in Fig.\  \ref{fig:ThinCrystals}a, quantum corrections affect the electron spectra across almost the entire energy range, including its peak.

From the discussion above we see that including the LL force for thin crystals has no effect on the radiation spectrum in the planar regime and almost no effect in the axial regime. Since the spin contribution is non-negligible for the axial case, we will use the BCK model in both the CES and stochastic schemes when calculating electron spectra.

Figure \ref{fig:models_compare}a displays the theoretical enhancement spectra for 50 GeV positrons traversing a 6.2 mm Si crystal, parallel to the (110) plane. Particle entry angles are confined between $\pm 30$ $\mu$rad with respect to the plane, and sampled from a Gaussian distribution with a mean of 0 and divergence of 100 $\mu$rad. The spectra are evaluated using each model and scheme described above, and it is evident that quantum effects are non-negligible even at $\chi = 0.015$.
We see that the substitution and BCK models (CES) give identical spectra around the peak, and that the BCK model includes the high-energy bremsstrahlung tail from random scattering in the crystal. This shows that quantum effects related to spin are negligible when looking at the channeling radiation, and only the kinematic effect of photon recoil is important in our regime. We also see that quantum stochastic effects are negligible. Therefore, we choose to use the substitution model in our analysis of the planar channeling spectra due to its simple kinematic interpretation.  

Figure \ref{fig:models_compare}b shows the theoretical spectra for 80 GeV electrons traversing a 1.0 mm diamond single crystal in the $\langle 100 \rangle$ axial channeling regime with incident angles $\psi < \psi_1$. The incoming
electrons have the same distribution as the positrons do in the planar case. It is clear that quantum effects are non-negligible.  There is a noticeable difference between the spectra produced using the LL equation with and without the damping factor $G(\chi)$. When the damping factor is included, the BCK model using the LL equation in the CES agrees quite well with the stochastic model in the region beyond the peak, but is is clear that stochastic effects cannot be neglected entirely in the axial channeling regime.

\begin{figure*}[ht!]
	\includegraphics[width=\textwidth]{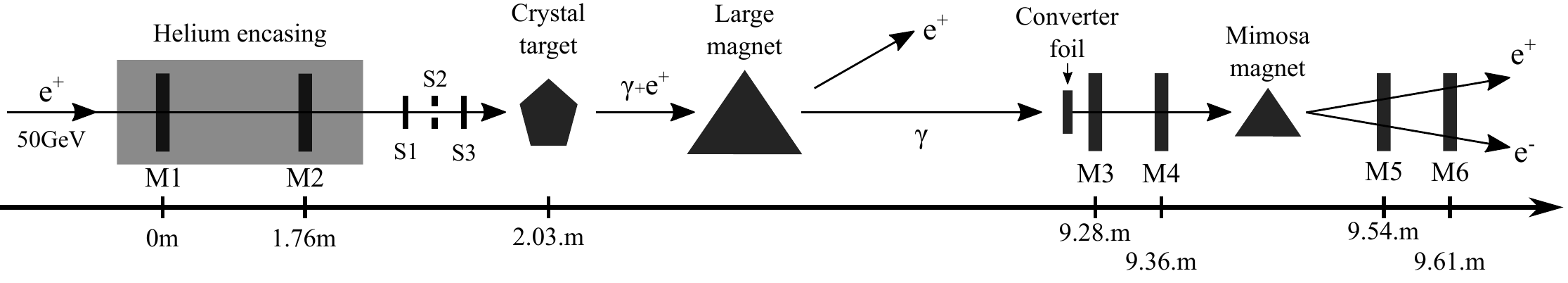}
	\caption{Experimental setup during the 2017 experiment. A schematic representation of the experimental setup in the H4 beam line in the SPS North Area at CERN. The symbols ``Sj'', with $j=1,2,3$, denote the scintillators and the symbols ``Mi'', with $i=1,\ldots,6$, denote `Mimosa-26' position-sensitive detectors with a resolution of $3.5$ $\mu$m.\label{fig:setupfig}}
\end{figure*}

\section{Experiments}
The experiments were performed at the H4 beamline of the CERN SPS by the NA63 collaboration. Four silicon single crystals and two diamond single crystals with thicknesses ranging from $1.0\;\mbox{mm}$ to $6.2\;\mbox{mm}$ were used, aligned with the beam along the $(110)$ plane or $\langle100\rangle$ axis. The experimental runs using silicon single crystals were performed in 2017, and the runs using diamond single crystals were performed in 2018. A schematic of the 2017 setup is shown in Fig.\ \ref{fig:setupfig}. Essentially the same setup was used for the 2018 experiment except the scintillators were placed between the crystal and the large magnet instead of between the helium encasing and the crystal. 

    A 200 $\mu$m converter foil of amorphous tantalum corresponding to $5\%$ of the radiation length (3.8\% chance of pair production), was employed to generate e+/e- pairs from the emitted photons which were subsequently  analyzed in a magnetic spectrometer, see Fig.\ \ref{fig:setupfig}. The magnetic spectrometer measures the deflection angle of both the electron and the positron produced by the photon, and by knowing the field of the magnet we can find the energy of each particle, and in turn the energy of the original photon, assuming that the energy of the converted photon is much larger than the rest mass of an electron.
    
Each Mimosa has a detection area of 1 cm $\times$ 2 cm, this means that alignment of the beam through the telescope arm spanned by M1--M6 has to be precise. It also means that we have an energy cutoff in the magnetic spectrometer, because low-energy particles produced by low-energy photons will be deflected outside the detection area of M5 and M6. Ensuring that low-energy photons are deflected outside M5 and M6 and by using a thin converter foil, this procedure enables us to measure the single-photon spectrum in the radiation-reaction regime where many photons are emitted by a single electron or positron, thus the problem of photon pile-up is avoided. 

The scintillators S1--S3 produce a trigger signal for the Mimosas to save their data buffer. S1 and S3 are mm thin scintillation sheets, while S2 is a cm thick scintillation sheet with a hole in its center with a radius of 5 mm. A successfull trigger is thus S1 + !S2 + S3, meaning that both S1 and S3 have to give a signal while S2 has to be quiet. As mentioned above, the placement of the scintillators has been switched between the 2017 experiment and the 2018 experiment. The selected locations each have advantages and disadvantages.  By placing the scintillators before the crystal, we are able to trigger on a much ``cleaner" beam, as it has not yet been perturbed by penetration of the crystal.  When placing the scintillators after the crystal, the initial direction of each particle determined by M1 and M2 has not yet been perturbed by the penetration of the scintillators before it hits the crystal. In practice we have not seen any difference between the two placements. 

For each beam configuration a background measurement has been made where the radiation is measured with the target crystal removed. The background consists of photons produced by bremsstrahlung from the beam hitting material such as vacuum windows, scintillators, collimators, and air gaps upstream of the large magnet. This background spectrum is subtracted from each measurement with a target resulting in a pure crystal spectrum. To mimic the experimental conditions, a simulation of the background spectrum is made by placing material in the beam path matching that which is present in the experiment. Simulated spectra with the target crystal in place then includes the material producing the background photons, and from those we subtract the simulated background with no crystal installed to obtain the pure crystal spectra.
The simulated background agrees well with the background measured in the experiment.

\section{Results and discussion}
\label{sec:results}
When the beam divergence is significantly larger than the critical Lindhard angle, as is the case in our experiments, many particles in the beam are initially outside the channeling region.  As a result, the average value of the strong-field parameter, $\bar{\chi}$, that the beam particles are exposed to, will be different from that experienced if they were all immediately channeled. For positrons in the planar channeling regime, $\bar{\chi}$ will be larger for the full beam than for channeled positrons. For electrons incident on a crystal oriented along an axis, $\bar{\chi}$ will be smaller for the full beam of electrons than for channeled electrons. Simulated values of $\bar{\chi}$, $\sqrt{\overline{\chi^2}}$ and expected energy loss $\Delta$ from \cref{eq:LLloss} for different energies, crystals, and entry angles are listed in Table \ref{tb:table2}.  

In the analysis of the experimental data and in our simulations we make various cuts on the angle of incidence of the projectiles in order to vary those parameters that are essential for our investigation. Among these, the strong-field parameter, defined in Eq. (\ref{eq:chi}), plays a dual role. In itself the magnitude of $\chi$ determines how strongly the spectra are influenced by quantum effects like recoil and spin. Entering the ratio, $\eta$, of the radiative damping force to the external force as $\eta =\gamma\chi\alpha$, Eq. (\ref{eq:eta}), the magnitude of $\chi$ further reflects the magnitude of the radiation reaction at given projectile energy. We wish to minimize quantum effects but aim to maximize the radiation reaction. Selecting particles with different impact angles, for instance aiming for those initially channeled versus excluding such particles, is one way of trying to learn how best to handle this conflict of interests. Another parameter we vary by making specific selections in angle of incidence is the ratio of the opening angle of the radiation cone to the typical angle of excursion of the projectiles whose motion in first approximation is governed by the continuum crystal potential. This parameter, which typically varies much more between the cuts than the strong-field parameter, is a measure of how well a constant-field approximation for the radiation-emission process will work.

\begin{figure*}[h]
	\begin{center}
		\includegraphics[width=1\textwidth]{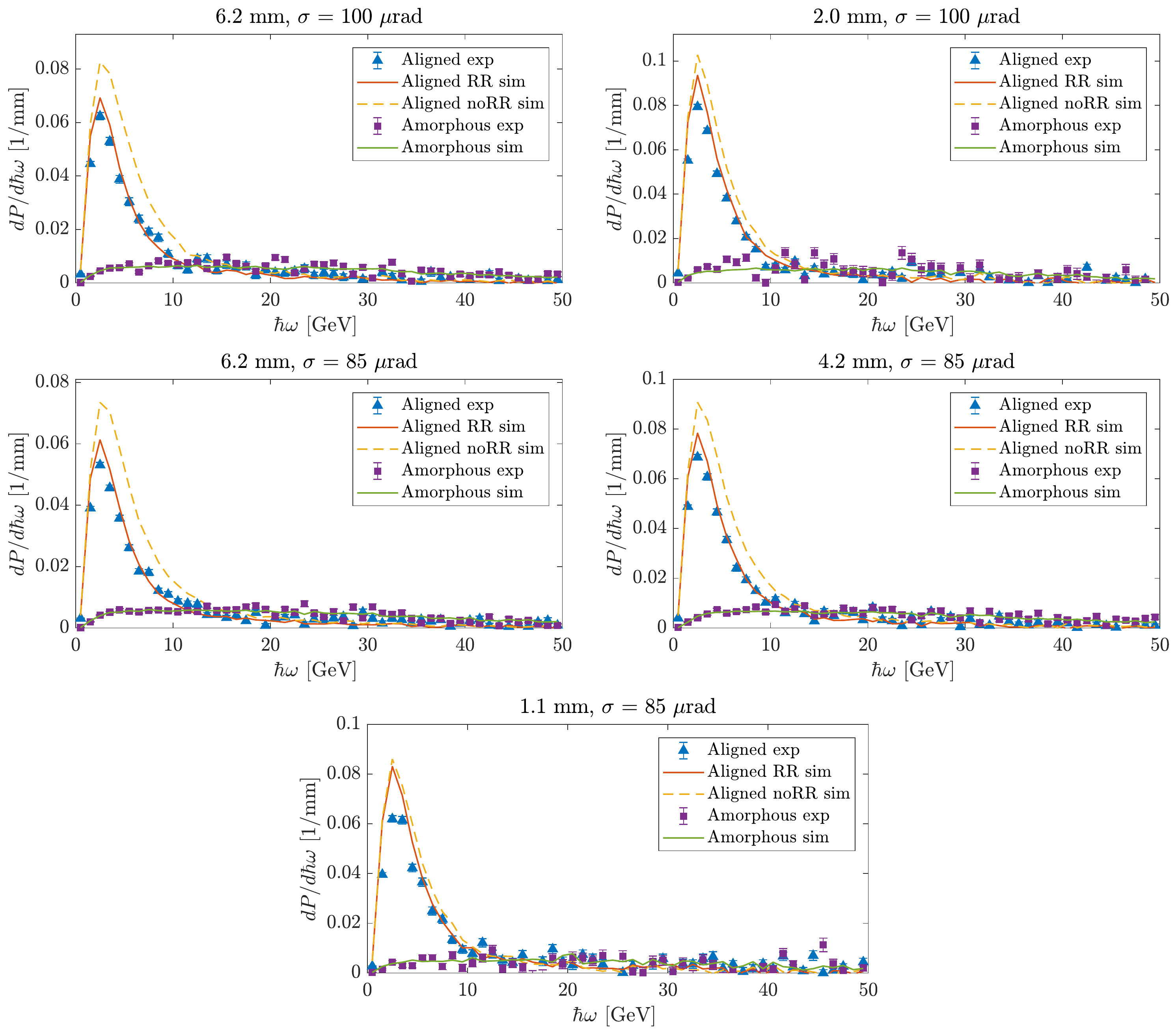} 
		\caption{Radiation power spectra obtained for 50 GeV positrons passing 1.1, 2.0, 4.2, and 6.2 mm thick silicon crystals aligned with the (110) plane, and the corresponding amorphous spectra. Only particles with entry angle between $\pm$ 30 $\mu$rad with respect to the crystal planes are included. The two top-most figures show experimental data and calculations obtained with a beam with a divergence of $\sigma_\perp=100~\mu$rad in the direction transverse to the plane, while the three remaining figures are obtained for a beam with a divergence of $\sigma_\perp=85~\mu$rad.	The theoretical spectra calculated using the substitution model in the CES are shown for trajectories deriving from the LL equation (``RR sim'') as red solid lines and for Lorentz-force trajectories (``noRR sim'') as yellow dashed lines. The simulated amorphous spectra (``Amorphous sim'') is shown in solid green curves. The data from the planar aligned crystal (``Aligned exp'') is shown in blue triangles while the data from the amorphous setting (``Amorphous exp'') appear as purple squares.}
		\label{fig:alignresults_pow_cut}
	\end{center}
\end{figure*}

\begin{figure*}[h]
	\begin{center}
		\includegraphics[width=1\textwidth]{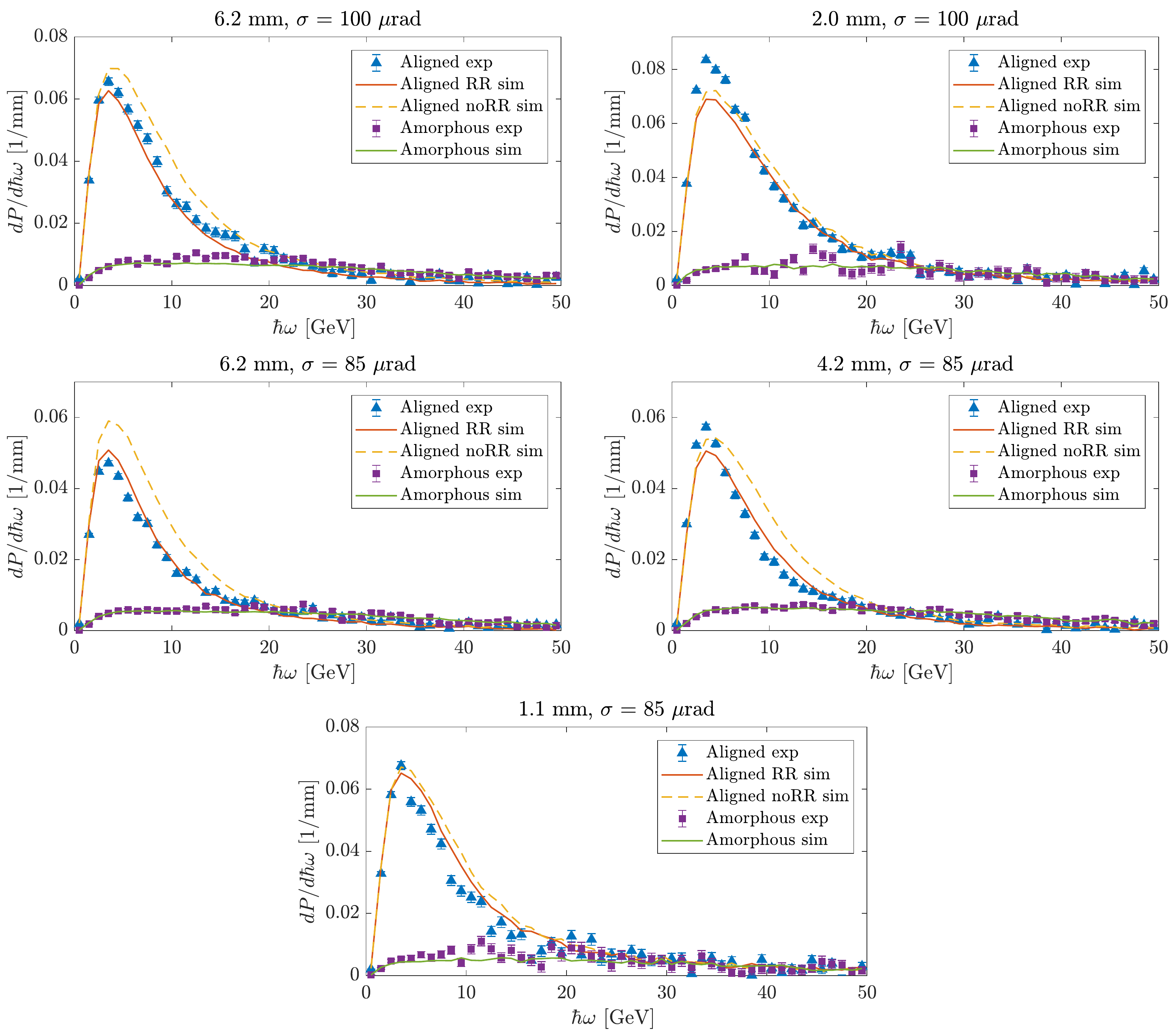} 
		\caption{As Figure \ref{fig:alignresults_pow_cut} but including all positrons in the beam.}
		\label{fig:alignresults_pow_full}
	\end{center}
\end{figure*}

\begin{figure*}[h]
	\begin{center} 
		\includegraphics[width=1\textwidth]{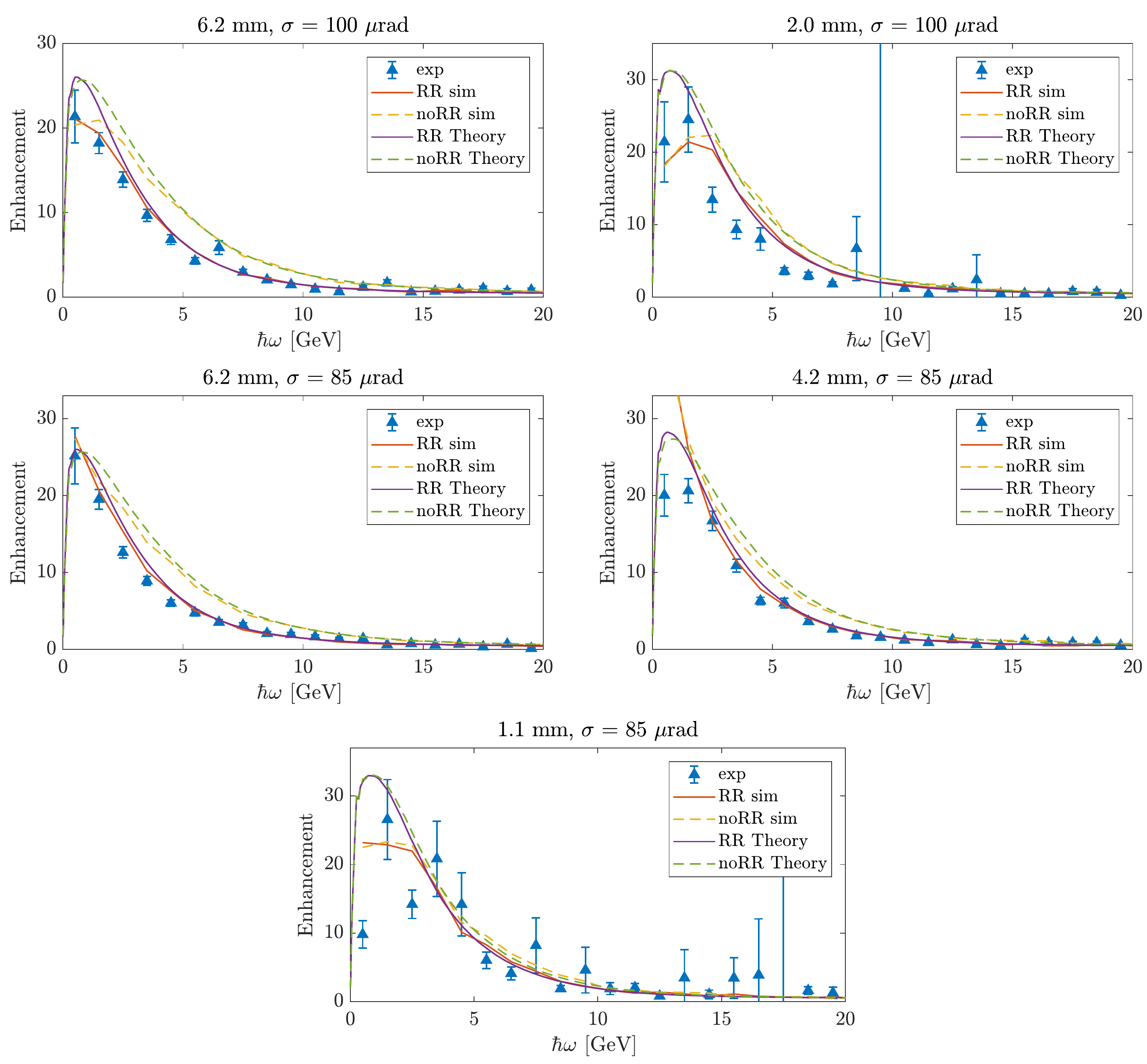} 
		\caption{Enhancement spectra for 50 GeV positrons passing 1.1, 2.0, 4.2, and 6.2 mm thick silicon crystals aligned to the (110) plane. Only particles with entry angle between $\pm$ 30 $\mu$rad with respect to the crystal planes are included. Solid lines pertain to spectra where radiation-reaction effects are included via the substitution model in the CES (``RR''). Dashed lines pertain to spectra where radiation-reaction effects are excluded (``noRR''). Triangles show experimental data with statistical error bars (``exp''). The purple and green lines show pure theoretical calculations of the channeling radiation divided by the analytical Bethe-Heitler bremsstrahlung spectrum (``Theory''), while the red and yellow curves display enhancement spectra where both the amorphous and the channeling spectrum have been through the experimental simulation and analysis routine (``sim''). }
		\label{fig:alignresults_enh_cut}
	\end{center}
\end{figure*}

\begin{figure*}[h]
	\begin{center} 
		\includegraphics[width=1\textwidth]{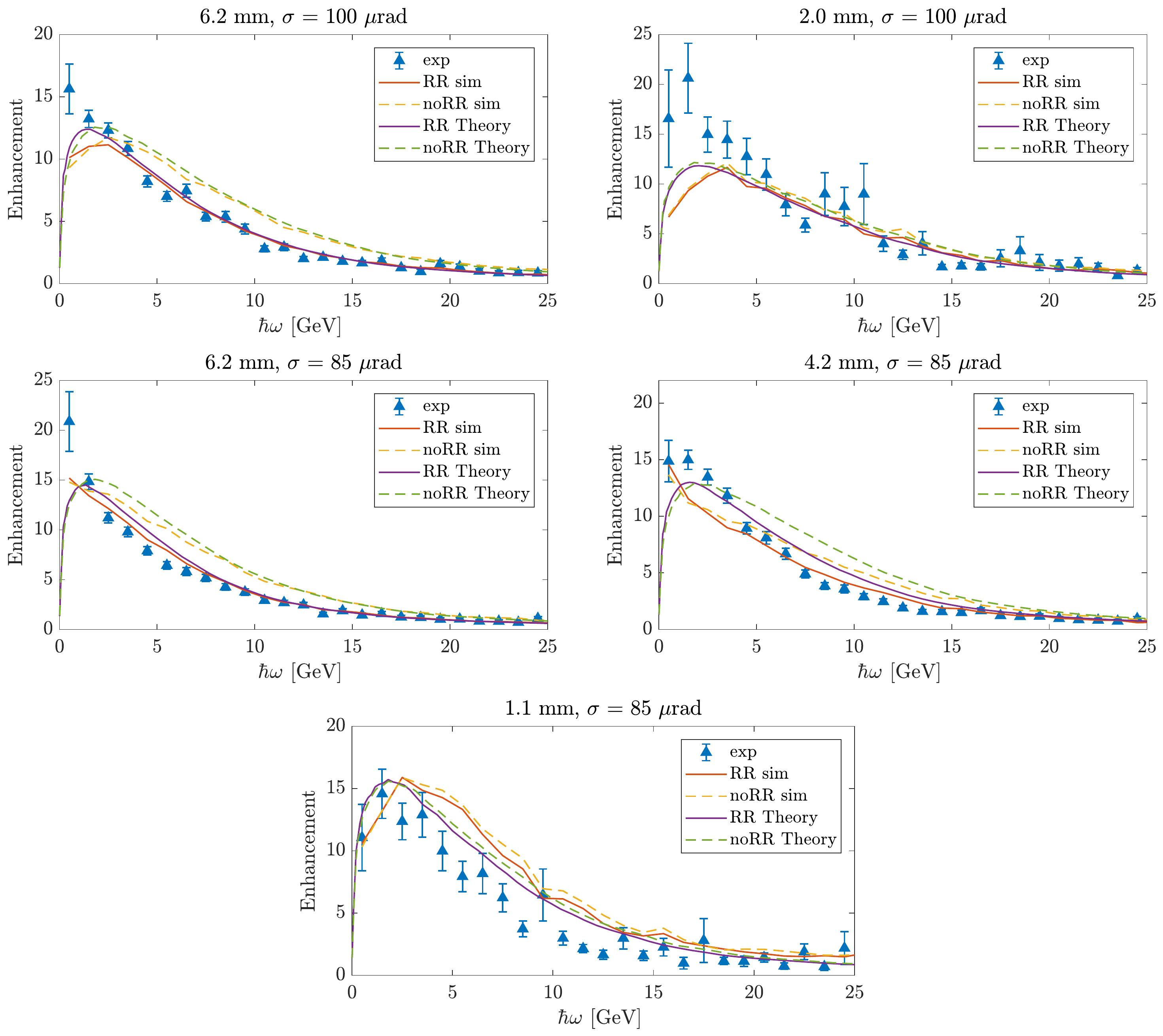} 
		\caption{As Figure \ref{fig:alignresults_enh_cut} but including all positrons in the beam.}
		\label{fig:alignresults_enh_full}
	\end{center}
\end{figure*}

For the positrons, an angular cut selecting particles with angles smaller than $30~\mu$rad to the planar direction is made. With this cut, which covers the channeling region (see Table \ref{tb:Lindhard_angle}), the maximum value of $\bar{\chi}$ is reached for the thinnest crystal, 1.1 mm, which is $\bar{\chi}\approx 0.014$, otherwise it is $\bar{\chi}\approx 0.016$ when no cuts are made.  As a result, the positrons are closer to the classical regime $\bar{\chi}\ll1$ than in the previous investigation \cite{Wistisen_2018} that was performed in the regime where quantum stochasticity, spin, and single photon recoil played a dominant role. We note that the value of $\bar{\chi}$ falls with increasing thickness of the crystals because the energy of the particle is lower in the later part of a thick crystal than in the beginning. Data and simulations for positrons using the $30~\mu$rad cut are shown in Figs.\ \ref{fig:alignresults_pow_cut} and  \ref{fig:alignresults_enh_cut}, while Figs.\ \ref{fig:alignresults_pow_full} and  \ref{fig:alignresults_enh_full} are without angular cuts. 

\begin{figure*}[h]
	\begin{center}
		\includegraphics[width=1\textwidth]{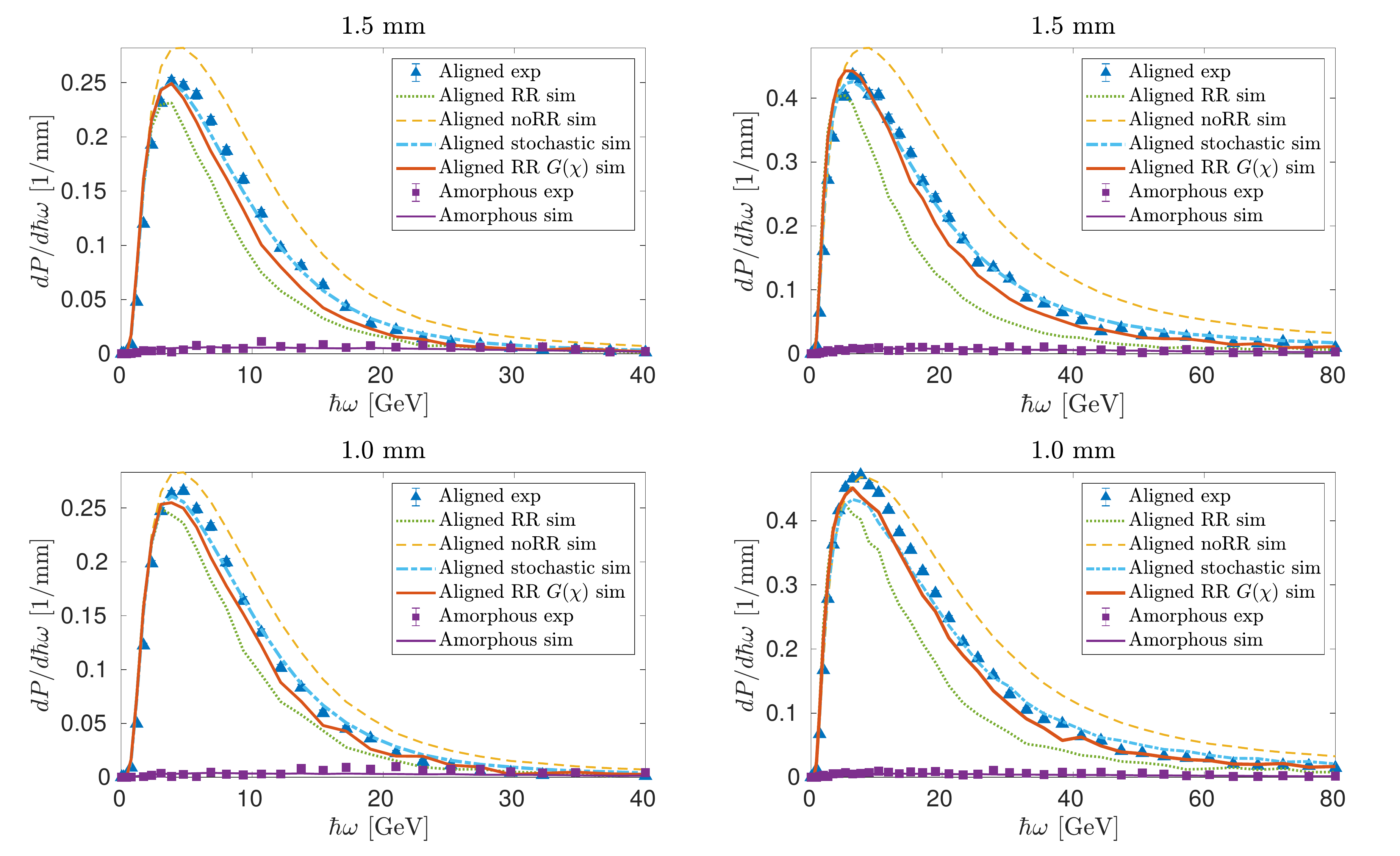} 
		\caption{Radiation power spectra obtained for 40 GeV (left) and 80 GeV (right) electrons traversing 1.0 mm (bottom) and 1.5 mm (top) thick diamond crystals aligned to the $\langle100\rangle$ axis, and the corresponding amorphous spectra. The beam divergences are $[\sigma_x,\sigma_y] =[192,89] $ $\mu$rad  and $[\sigma_x,\sigma_y] =[129,75] $ $\mu$rad for the 40 GeV and 80 GeV beam, respectively, with $x$ and $y$ both approximately aligned with (110) planes. The theoretical spectra calculated using the BCK model in the CES are shown for trajectories deriving from the LL equation without the $G(\chi)$ correction (``RR sim'') as green dotted lines, from the LL equation including the $G(\chi)$ correction (``RR $G(\chi)$ sim'') as red solid lines, and for pure Lorentz-force trajectories (``noRR sim'') as yellow dashed lines. The spectra calculated using the BCK model in the stochastic scheme (``stochastic sim'') are shown as dashed-dotted blue lines while the simulated amorphous spectra (``Amorphous sim'') appear solid purple. The data from the axially aligned crystal (``Aligned exp'') is shown in blue triangles while the data from the amorphous setting (``Amorphous exp'') appear in purple squares.}
		\label{fig:alignresults_2018_AlignedNoCuts}
	\end{center}
\end{figure*}

\begin{figure*}[h]
	\begin{center}
		\includegraphics[width=1\textwidth]{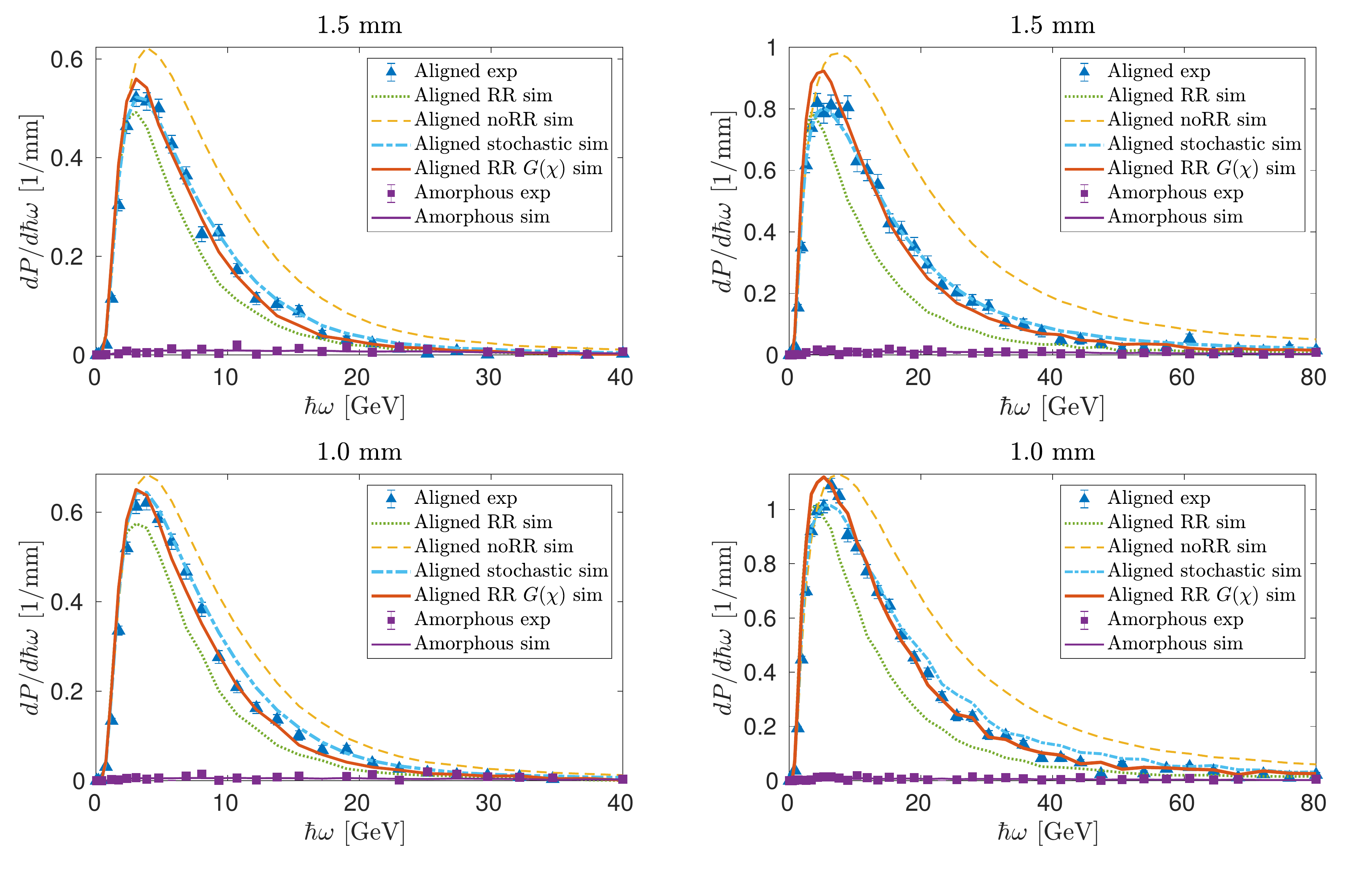} 
		\caption{As Figure \ref{fig:alignresults_2018_AlignedNoCuts} but including only electrons with entry angle less than the critical Lindhard angle $\psi_1$ with respect to the crystal axis.}
		\label{fig:alignresults_2018_CriticalAngleCuts}
	\end{center}
\end{figure*}

\begin{figure*}[h]
	\begin{center}
		\includegraphics[width=1\textwidth]{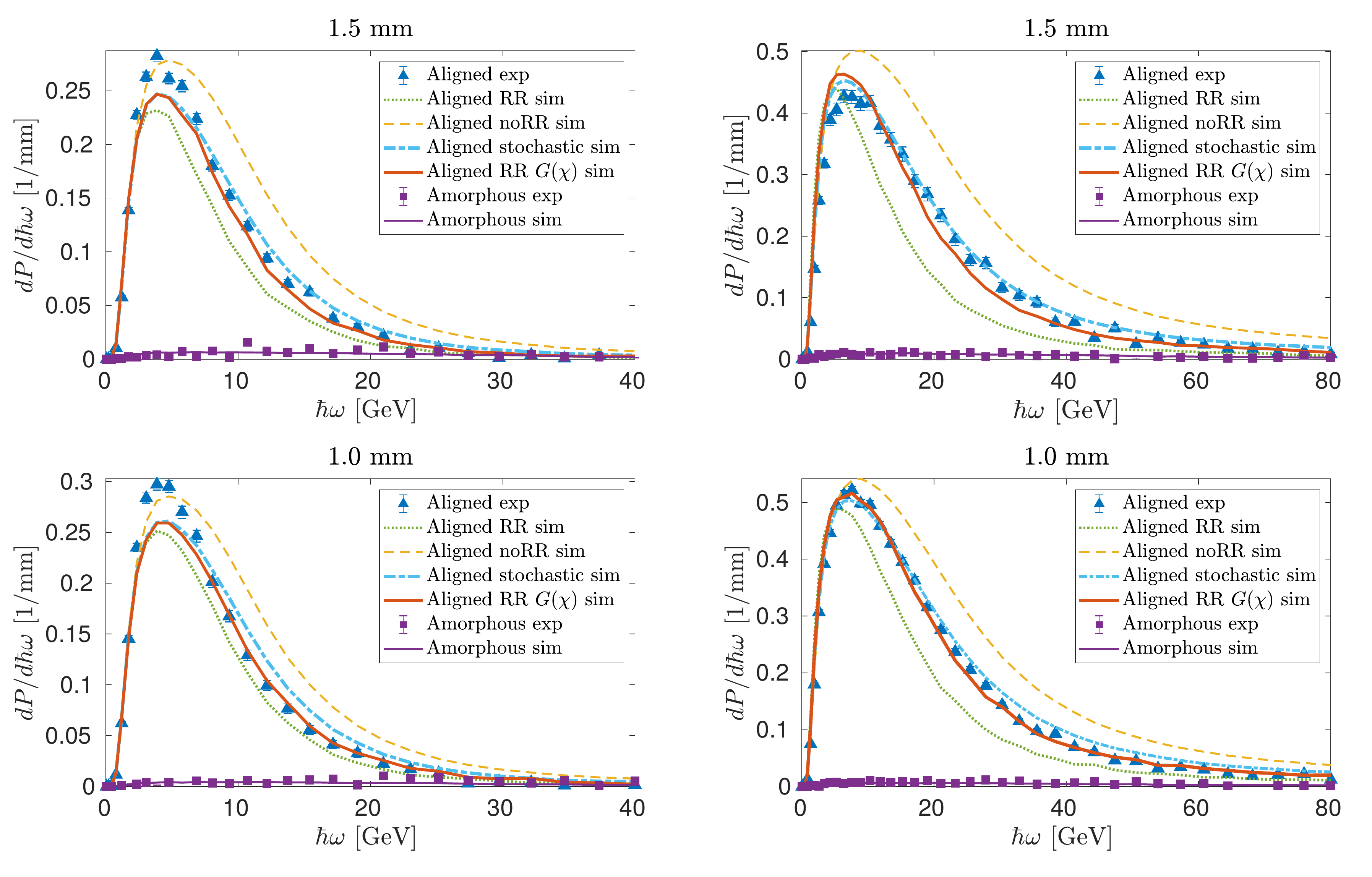} 
		\caption{As Figure \ref{fig:alignresults_2018_AlignedNoCuts} but including only electrons with entry angle between $2\psi_1$ and $4\psi_1$ with respect to the crystal axis.}
		\label{fig:alignresults_2018_DonutCuts}
	\end{center}
\end{figure*}

For the electrons we have made two angular cuts. The first cut selects particles with angles smaller than the critical angle (see Table \ref{tb:Lindhard_angle}) and hence includes all initially channeled electrons. The second cut selects particles in a ``donut'' around the axis, with angles between two and four times the critical angle, and thereby excludes all channeled electrons. For both cuts we have $\bar{\chi}\approx 0.06$ for the 1.0 mm crystal at 80 GeV, with the value for the donut falling 14 \% below that for the central cut. Such value of the strong-field parameter is large enough that quantum effects significantly influence the spectrum but, yet, it is smaller than that used previously  \cite{Wistisen_2018}. Spectra for the critical-angle cut are shown in Fig.\ \ref{fig:alignresults_2018_CriticalAngleCuts} and spectra for the ``donut''-cut appear in Fig.\  \ref{fig:alignresults_2018_DonutCuts}. Figure \ref{fig:alignresults_2018_AlignedNoCuts} displays spectra for electrons without any cuts on the angle of incidence.

As seen in \cref{tb:table2}, \cref{eq:LLloss} substantially overestimates the energy loss compared to the simulated energy loss for the electrons when including the $G(\chi)$ correction. On the other hand, when excluding $G(\chi)$ everywhere, $\Delta$  is never more than 1\% off $\Delta E_{LL}$ (not shown in the table). For the positrons the prediction underestimates the energy loss by around $20\%$ in all cases compared to the simulated results without the $G(\chi)$ correction. As the simulation includes energy loss by incoherent scattering based on \cref{eq:ElossMSc}, we expect that the simulated energy loss would be higher than what \cref{eq:LLloss} predicts. The ratio of energy lost due to incoherent scattering and the LL equation is much lower for the electrons than for the positrons where this ratio is around $10\%-20\%$ and only a few percent for the electrons, explaining the underestimated energy loss by \cref{eq:LLloss}.

``Amorphous spectra'' are recorded and simulated for both positrons and electrons by orienting the crystals far from all low-index crystallographic orientations. These spectra are in good agreement with the Bethe-Heitler spectrum for amorphous targets of the same material and density, when a combined detection efficiency of about $3/4$ (with slight variations between targets due to the small variation in the beam conditions) is taken into account. Since the Bethe-Heitler spectrum is well-known, this constitutes a reassuring test of the simulation algorithm.

Having experimental data for both the axial and planar channeling regimes provides us with the unique opportunity to investigate the applicability of the LL equation over a wide range of values of $\chi$ and to estimate the regions where quantum effects can be treated as perturbations to the classical theory.

Figure \ref{fig:alignresults_pow_cut} shows power spectra for a 30 $\mu$rad angular cut on the incident positron beam. The power spectrum is the intensity spectrum divided by the crystal thickness, hence the 1/mm dimension. A clear agreement is observed between the experimental data and theoretical spectra based on the LL equation and the substitution model for all crystal thicknesses. When excluding the radiation-reaction force contained in the LL equation, the discrepancy between data and theory increases with the crystal thickness. For entry angles less than 30 $\mu$rad the coherent part of the spectrum originating from the motion in continuum potential reaches up to 10--15 GeV. Radiation beyond this point is due to incoherent scattering, mainly multiple Coulomb scattering on nuclei. For well-channeled positrons moving between a set of planes, the average density of nuclei encountered along their trajectories is much lower than in an amorphous material. This means that the incoherent part of the aligned spectrum is lower than the amorphous spectrum.

The positron data are taken with two beam configurations with different angular divergence $\sigma$ and beam direction with respect to the detectors, due to an unexpected long-duration interruption of the SPS caused by technical problems. The beam with the smaller divergence, $\sigma = 85$ $\mu$rad, has an entry angle which is slightly tilted with respect to the telescope arm spanned by M1--M6, meaning that less particles will reach M3--M6, but the beam hits the scintillator hole with the same amount of particles as with the $\sigma=100$ $\mu$rad beam. For this reason the tighter beam has a lower intensity than the broader beam. 

Figure \ref{fig:alignresults_pow_full} displays the power spectra for the full positron beam. The peak intensity is essentially as for the cut, Fig.\  \ref{fig:alignresults_pow_cut}, but due to the higher value of $\bar{\chi}$, the coherent part of the spectra extends to higher photon energies.   As is particularly evident for the thicker crystals, the experimental data again clearly favors the simulation based on the LL equation over that accounting for the pure Lorentz force only.  Since all particles in the beam are included with a beam divergence reaching $\sigma = 100$ $\mu$rad, particles in the angular distribution with large angles are likely not to reach M3--M6. Under this condition it is challenging to simulate the experiment because the position of the detectors and the beam entry angles are input parameters of the simulation. As a result, the discrepancies seen between data and simulation spectra generally are larger for the full beam compared to the spectra in Fig.\ \ref{fig:alignresults_pow_cut} where angular cuts have been applied. Nevertheless it is clear that inclusion of the LL damping force is essential. With the full beam we do not see an effect of beam focusing on the incoherent part of the spectrum since only a minor fraction of the positrons are channeled. 

Figure \ref{fig:alignresults_2018_AlignedNoCuts} shows power spectra for the full electron beam.  For electrons, the difference between the theoretical curves including and excluding the radiation reaction as described by the LL equation are large even for the 1 mm crystal.  The experimental data falls roughly half way between the two sets of theory curves in all cases.  For the full electron beam conditions (Table \ref{tb:table2}), $\bar{\chi}$ varies from ~0.03 to 0.06.  At such values of $\bar{\chi}$, even though smaller than 1, the LL equation overestimates the radiation emitted leaving the electrons with less energy going forward through the crystal.  This results in lowering the emission rate for the remaining length of the crystal and an underestimation of the total radiation emitted. The conditions are reflected by the damping factor $G(\chi)$ attaining values considerably less than 1, for instance, $G(0.06)=0.76$. When multiplying $G(\chi)$ on the radiative damping force in the LL equation the resulting radiation spectra fit the experimental findings quite well but slightly underestimate the data. The stochastic scheme, which is not based on the LL equation, also produces spectra that fits the experimental data well with, in general, even less deviations.

By comparing the electron and the positron data, it is evident that electrons at lower energy on axis produce a much higher enhancement over the amorphous spectrum than do positrons at a higher energy on a plane. One reason is that axial continuum potentials are stronger than their planar counterparts by a factor of up to about 10, another is that electrons spend more time than positrons in the vicinity of the nuclei which is the location of the strongest electric field. Typically the total radiative energy loss is higher for the electrons than for the positrons for the crystals used in our experiments, cf. Table \ref{tb:table2}.   

Figure\ \ref{fig:alignresults_2018_CriticalAngleCuts} displays power spectra for electrons excluding ones incident to the crystal axis at angles larger than the Lindhard critical angle $\psi_1$; see Table \ref{tb:Lindhard_angle} for actual values of $\psi_1$.  With this cut, the radiation process for electrons following trajectories determined by the continuum potential proceeds as if they are in a locally constant electromagnetic field.  This is especially true for the highest electron energy, cf. Subsec.\ \ref{subsec:CFA}. Accordingly, the simulated spectra, based on the LL equation with the $G$-factor included on the damping term, fit the experimental data very well. Generally, these spectra reproduce the data as well as does the stochastic scheme which underlines the effectiveness of the LL equation when the quantum effects in the radiation process are properly included. Only near the peak in some of the spectra does the stochastic scheme appear slightly superior.  

Applying the cuts in Fig.\ \ref{fig:alignresults_2018_CriticalAngleCuts} to the incident electron beam we see that more photons are emitted on average per electron than for the full beam but also that the hardness of the spectrum is lower than in the full-beam spectra. By comparing Figs.\ \ref{fig:alignresults_2018_AlignedNoCuts} and \ref{fig:alignresults_2018_CriticalAngleCuts} we observe that confinement of particles to the axis enhances the spectrum more at lower than at higher energies, with the high-energy end of the spectrum essentially remaining the same up to focussing effects due to channeling.  

Figure \ref{fig:alignresults_2018_DonutCuts} shows power spectra where the incoming electrons are confined to entry angles between $2\psi_1$ and $4\psi_1$ with respect to the crystal axis. These spectra are similar to the spectra obtained with the full beam, Fig.\ \ref{fig:alignresults_2018_AlignedNoCuts}.  Since much of the beam lies within this cut, the similarity between spectra is no surprise. It is worth noting that the agreement between the stochastic curve and the simulation based on LL equation, including the damping factor $G(\chi)$, is nearly perfect around the peak. Yet, when electrons move well outside the channeling region, the procedure of applying the damping factor is less justified as $G$ pertains to emission in a constant field.  

The photon emission spectra, shown in Figures\ \ref{fig:alignresults_pow_cut} to \ref{fig:alignresults_2018_DonutCuts}, display features that cannot be reproduced if only the Lorentz force is used to calculate the particle trajectories. Moreover, the agreement is remarkably good between our experimental data and theory that includes radiation reaction according to the LL equation with proper inclusion of quantum effects.  Nevertheless, due to the unavoidable quantum corrections introduced in the radiation spectra, there exists a discrepancy between the energy loss predicted by the LL equation and the energy loss obtained by integrating the radiation spectrum. In the planar channeling regime, this discrepancy is 16--17\% for well-channeled particles and approximately 30\% for the full positron beam (see Table \ref{tb:table2}). The discrepancy easily reaches 50\% for the electrons in our experiment. The discrepancy is smallest for well-channeled positrons since they emit the relatively softest radiation whereby a smaller quantum correction results.  Non-channeled positrons and the electrons have a larger discrepancy. The discrepancy is generally largest in the axial channeling regime due to the stronger fields. As is evident from Table \ref{tb:table2}, inclusion of the damping factor $G(\chi)$ improves the situation dramatically by producing a substantial reduction of discrepancies.
The modest energy loss difference of 7\% for $\psi<\psi_1$ when including the damping factor at the highest electron energy, can be taken as an indication of the quality of the procedure applied and validates the constant-field approach to the radiation process for high-energy channeled electrons.  

The agreement between experiment and theory-based simulation where we include the LL equation is supported by looking at the enhancement spectra on Fig.\ \ref{fig:alignresults_enh_cut} and Fig.\ \ref{fig:alignresults_enh_full}.
Here the theory-based simulation curves do \emph{not} rely on an intricate analysis algorithm as we divide the theoretical curves by the analytical Bethe-Heitler spectrum. These enhancement spectra are directly comparable to the experimental enhancement spectra as the perturbations to the spectra from the experimental setup is removed during division. 
In addition, the experimental data is directly based on data obtained from the aligned crystal divided by data obtained in the amorphous orientation (Bethe-Heitler), and are thus, at least to first order, independent of selection criteria for the pairs, detection efficiencies etc. Due to poor statistics of the amorphous data set for the diamond single crystals we do not show similar figures for the axial case but since the simulation includes the same physical effects the same conclusion can be drawn.


\section{Conclusion}
\label{sec:conclusion}
We investigated the necessity of radiation reaction for several unique cases near the classical limit in aligned crystals. The energy losses are moderate in the planar channeling regime for 50 GeV positrons so it could be expected that the spectra obtained neglecting radiation-reaction effects would be roughly adequate, but the experiments and simulations clearly show that this is not the case. The results of our experiments demonstrate that the pure Lorentz force is inadequate to describe the dynamics of electric charges when moving in strong electromagnetic fields. The simulations without radiation reaction overestimate the emitted radiation for both axial and planar channeling, even with $\bar{\chi} \ll 1$. In contrast, predictions according to the Landau-Lifshitz equation, when accounting for the photon recoil in the radiation spectra through the substitution model, result in remarkable agreement with the experimental radiation spectra pertaining to planar channeled 50 GeV positrons. For the axially channeled electrons in our experiment it is even more essential to include the radiation reaction than for the positrons.  For axially channeled 40 GeV and 80 GeV electrons we find that theoretical spectra calculated using trajectories obtained from the Landau-Lifshitz equation with the radiation-reaction force multiplied by $G(\chi)$ convincingly reproduce our experimental data for all cuts, energies, and crystal thicknesses when accounting for photon recoil and particle spin. 
Due to the large energy loss experienced by the particle as it traverses the crystal, exclusion of the $G(\chi)$-correction leads to an underestimation of the emitted radiation intensity in the latter parts of the crystal. We note further that the stochastic scheme based on the BCK model generally reproduces the theoretical spectra well and, as noted also in \cite{cern2017}, that this scheme does not rely on the constant-field approximation which is the origin of the $G(\chi)$ factor.

Recording the radiation emitted at different values of the strong-field parameter, $\bar{\chi} \approx$ 0.01 -- 0.06 (see Table \ref{tb:table2}), has allowed us to probe the border between the classical and the quantum-mechanical description of radiation reaction. We find that particularly planar channeling of 50 GeV positrons provide a regime where quantum corrections cannot be neglected but they also do not dominate, a compromise that must be made to illustrate the effect of the LL equation. We further see that even with a relatively small value of the strong-field parameter, as $\chi \simeq 0.03$ encountered by 40 GeV electrons in the axial channeling regime of a diamond crystal aligned along the $\langle 100\rangle$ axis, quantum effects (particle spin, photon recoil, and reduction of the radiation intensity) have a significant influence on the radiation-emission spectrum. The most drastic influence of radiation reaction encountered in our experiments, and of quantum effects, appears for axially channeled electrons at the highest applied energy, 80 GeV, where $\chi$ attains its largest value $\chi \simeq 0.06$.

The detailed results and analysis presented above clearly demonstrate that the Landau-Lifshitz equation, with suitable modifications, is applicable in a very wide regime of combinations of particle energies, crystal orientations, materials, and thicknesses to describe the phenomenon of radiation reaction. Overall, our results strongly point towards the LL equation as a very satisfactory answer to the century-old problem of radiation reaction in classical electrodynamics.\\
\\

\begin{acknowledgements}
We are grateful for the assistance provided by T.N. Wistisen in designing and executing the experiments.
We further acknowledge the superb technical help and expertise from Per Bluhme Christensen,
Erik Loft Larsen and Frank Daugaard (AU) in setting up the experiment
and data acquisition. The numerical results presented in this work were partly obtained at the Centre for Scientific Computing Aarhus (CSCAA) and with support from Nvidia's GPU grant program. This work was partially supported by the U.S. National Science Foundation (Grant No. PHY-1535696). We acknowledge the funding from nice https://nice.ku.dk/.
\end{acknowledgements}

%\bibliography{references}

%merlin.mbs apsrev4-1.bst 2010-07-25 4.21a (PWD, AO, DPC) hacked
%Control: key (0)
%Control: author (72) initials jnrlst
%Control: editor formatted (1) identically to author
%Control: production of article title (-1) disabled
%Control: page (0) single
%Control: year (1) truncated
%Control: production of eprint (0) enabled
\begin{thebibliography}{0}%
\makeatletter
\providecommand \@ifxundefined [1]{%
 \@ifx{#1\undefined}
}%
\providecommand \@ifnum [1]{%
 \ifnum #1\expandafter \@firstoftwo
 \else \expandafter \@secondoftwo
 \fi
}%
\providecommand \@ifx [1]{%
 \ifx #1\expandafter \@firstoftwo
 \else \expandafter \@secondoftwo
 \fi
}%
\providecommand \natexlab [1]{#1}%
\providecommand \enquote  [1]{``#1''}%
\providecommand \bibnamefont  [1]{#1}%
\providecommand \bibfnamefont [1]{#1}%
\providecommand \citenamefont [1]{#1}%
\providecommand \href@noop [0]{\@secondoftwo}%
\providecommand \href [0]{\begingroup \@sanitize@url \@href}%
\providecommand \@href[1]{\@@startlink{#1}\@@href}%
\providecommand \@@href[1]{\endgroup#1\@@endlink}%
\providecommand \@sanitize@url [0]{\catcode `\\12\catcode `\$12\catcode
  `\&12\catcode `\#12\catcode `\^12\catcode `\_12\catcode `\%12\relax}%
\providecommand \@@startlink[1]{}%
\providecommand \@@endlink[0]{}%
\providecommand \url  [0]{\begingroup\@sanitize@url \@url }%
\providecommand \@url [1]{\endgroup\@href {#1}{\urlprefix }}%
\providecommand \urlprefix  [0]{URL }%
\providecommand \Eprint [0]{\href }%
\providecommand \doibase [0]{http://dx.doi.org/}%
\providecommand \selectlanguage [0]{\@gobble}%
\providecommand \bibinfo  [0]{\@secondoftwo}%
\providecommand \bibfield  [0]{\@secondoftwo}%
\providecommand \translation [1]{[#1]}%
\providecommand \BibitemOpen [0]{}%
\providecommand \bibitemStop [0]{}%
\providecommand \bibitemNoStop [0]{.\EOS\space}%
\providecommand \EOS [0]{\spacefactor3000\relax}%
\providecommand \BibitemShut  [1]{\csname bibitem#1\endcsname}%
\let\auto@bib@innerbib\@empty
%</preamble>
\end{thebibliography}%


\begin{thebibliography}{58}%
\makeatletter
\providecommand \@ifxundefined [1]{%
 \@ifx{#1\undefined}
}%
\providecommand \@ifnum [1]{%
 \ifnum #1\expandafter \@firstoftwo
 \else \expandafter \@secondoftwo
 \fi
}%
\providecommand \@ifx [1]{%
 \ifx #1\expandafter \@firstoftwo
 \else \expandafter \@secondoftwo
 \fi
}%
\providecommand \natexlab [1]{#1}%
\providecommand \enquote  [1]{``#1''}%
\providecommand \bibnamefont  [1]{#1}%
\providecommand \bibfnamefont [1]{#1}%
\providecommand \citenamefont [1]{#1}%
\providecommand \href@noop [0]{\@secondoftwo}%
\providecommand \href [0]{\begingroup \@sanitize@url \@href}%
\providecommand \@href[1]{\@@startlink{#1}\@@href}%
\providecommand \@@href[1]{\endgroup#1\@@endlink}%
\providecommand \@sanitize@url [0]{\catcode `\\12\catcode `\$12\catcode
  `\&12\catcode `\#12\catcode `\^12\catcode `\_12\catcode `\%12\relax}%
\providecommand \@@startlink[1]{}%
\providecommand \@@endlink[0]{}%
\providecommand \url  [0]{\begingroup\@sanitize@url \@url }%
\providecommand \@url [1]{\endgroup\@href {#1}{\urlprefix }}%
\providecommand \urlprefix  [0]{URL }%
\providecommand \Eprint [0]{\href }%
\providecommand \doibase [0]{http://dx.doi.org/}%
\providecommand \selectlanguage [0]{\@gobble}%
\providecommand \bibinfo  [0]{\@secondoftwo}%
\providecommand \bibfield  [0]{\@secondoftwo}%
\providecommand \translation [1]{[#1]}%
\providecommand \BibitemOpen [0]{}%
\providecommand \bibitemStop [0]{}%
\providecommand \bibitemNoStop [0]{.\EOS\space}%
\providecommand \EOS [0]{\spacefactor3000\relax}%
\providecommand \BibitemShut  [1]{\csname bibitem#1\endcsname}%
\let\auto@bib@innerbib\@empty
%</preamble>
\bibitem [{\citenamefont {Jackson}(1975)}]{Jackson_b_1975}%
  \BibitemOpen
  \bibfield  {author} {\bibinfo {author} {\bibfnamefont {J.~D.}\ \bibnamefont
  {Jackson}},\ }\href@noop {} {\emph {\bibinfo {title} {Classical
  Electrodynamics}}}\ (\bibinfo  {publisher} {Wiley, New York},\ \bibinfo
  {year} {1975})\BibitemShut {NoStop}%
\bibitem [{\citenamefont {Landau}\ and\ \citenamefont
  {Lifshitz}(1975)}]{Landau_b_2_1975}%
  \BibitemOpen
  \bibfield  {author} {\bibinfo {author} {\bibfnamefont {L.~D.}\ \bibnamefont
  {Landau}}\ and\ \bibinfo {author} {\bibfnamefont {E.~M.}\ \bibnamefont
  {Lifshitz}},\ }\href@noop {} {\emph {\bibinfo {title} {The Classical Theory
  of Fields}}}\ (\bibinfo  {publisher} {Elsevier, Oxford},\ \bibinfo {year}
  {1975})\BibitemShut {NoStop}%
\bibitem [{\citenamefont {Abraham}(1905)}]{Abraham_b_1905}%
  \BibitemOpen
  \bibfield  {author} {\bibinfo {author} {\bibfnamefont {M.}~\bibnamefont
  {Abraham}},\ }\href@noop {} {\emph {\bibinfo {title} {Theorie der
  Elektrizit{\"a}t}}}\ (\bibinfo  {publisher} {Teubner, Leipzig},\ \bibinfo
  {year} {1905})\BibitemShut {NoStop}%
\bibitem [{\citenamefont {Lorentz}(1909)}]{Lorentz_b_1909}%
  \BibitemOpen
  \bibfield  {author} {\bibinfo {author} {\bibfnamefont {H.~A.}\ \bibnamefont
  {Lorentz}},\ }\href@noop {} {\emph {\bibinfo {title} {The Theory of
  Electrons}}}\ (\bibinfo  {publisher} {Teubner, Leipzig},\ \bibinfo {year}
  {1909})\BibitemShut {NoStop}%
\bibitem [{\citenamefont {Dirac}(1938)}]{Dirac_1938}%
  \BibitemOpen
  \bibfield  {author} {\bibinfo {author} {\bibfnamefont {P.~A.~M.}\
  \bibnamefont {Dirac}},\ }\href@noop {} {\bibfield  {journal} {\bibinfo
  {journal} {Proc. R. Soc. London, Ser. A}\ }\textbf {\bibinfo {volume}
  {167}},\ \bibinfo {pages} {148} (\bibinfo {year} {1938})}\BibitemShut
  {NoStop}%
\bibitem [{\citenamefont {Barut}(1980)}]{Barut_b_1980}%
  \BibitemOpen
  \bibfield  {author} {\bibinfo {author} {\bibfnamefont {A.}~\bibnamefont
  {Barut}},\ }\href@noop {} {\emph {\bibinfo {title} {Electrodynamics and
  Classical Theory of Fields and Particles}}}\ (\bibinfo  {publisher} {Dover
  Publications, New York},\ \bibinfo {year} {1980})\BibitemShut {NoStop}%
\bibitem [{\citenamefont {Rohrlich}(2007)}]{Rohrlich_b_2007}%
  \BibitemOpen
  \bibfield  {author} {\bibinfo {author} {\bibfnamefont {F.}~\bibnamefont
  {Rohrlich}},\ }\href@noop {} {\emph {\bibinfo {title} {Classical Charged
  Particles}}}\ (\bibinfo  {publisher} {World Scientific, Singapore},\ \bibinfo
  {year} {2007})\BibitemShut {NoStop}%
\bibitem [{\citenamefont {McDonald}(2018)}]{McDo18}%
  \BibitemOpen
  \bibfield  {author} {\bibinfo {author} {\bibfnamefont {K.~T.}\ \bibnamefont
  {McDonald}},\ }\href
  {http://www.hep.princeton.edu/~mcdonald/examples/selfforce.pdf} {\enquote
  {\bibinfo {title} {On the history of the radiation reaction},}\ } (\bibinfo
  {year} {2018}),\ \bibinfo {note} {notes, Princeton University}\BibitemShut
  {NoStop}%
\bibitem [{\citenamefont {Baier}\ \emph {et~al.}(1998)\citenamefont {Baier},
  \citenamefont {Katkov},\ and\ \citenamefont {Strakhovenko}}]{Baie98}%
  \BibitemOpen
  \bibfield  {author} {\bibinfo {author} {\bibfnamefont {V.~N.}\ \bibnamefont
  {Baier}}, \bibinfo {author} {\bibfnamefont {V.~M.}\ \bibnamefont {Katkov}}, \
  and\ \bibinfo {author} {\bibfnamefont {V.~M.}\ \bibnamefont {Strakhovenko}},\
  }\href@noop {} {\emph {\bibinfo {title} {{Electromagnetic processes at high
  energies in oriented single crystals}}}}\ (\bibinfo  {publisher} {Singapore,
  Singapore: World Scientific (1998) 554 p},\ \bibinfo {year}
  {1998})\BibitemShut {NoStop}%
%%CITATION = INSPIRE-487563;%%
\bibitem [{\citenamefont {Uggerh\o{}j}(2005)}]{Ugge05}%
  \BibitemOpen
  \bibfield  {author} {\bibinfo {author} {\bibfnamefont {U.~I.}\ \bibnamefont
  {Uggerh\o{}j}},\ }\href {https://link.aps.org/doi/10.1103/RevModPhys.77.1131}
  {\bibfield  {journal} {\bibinfo  {journal} {Rev. Mod. Phys.}\ }\textbf
  {\bibinfo {volume} {77}},\ \bibinfo {pages} {1131} (\bibinfo {year}
  {2005})}\BibitemShut {NoStop}%
\bibitem [{\citenamefont {S{\o}rensen}(1996)}]{Sorensen1996}%
  \BibitemOpen
  \bibfield  {author} {\bibinfo {author} {\bibfnamefont {A.~H.}\ \bibnamefont
  {S{\o}rensen}},\ }\href
  {http://www.sciencedirect.com/science/article/pii/0168583X96003497}
  {\bibfield  {journal} {\bibinfo  {journal} {Nuclear Instruments and Methods
  in Physics Research Section B: Beam Interactions with Materials and Atoms}\
  }\textbf {\bibinfo {volume} {119}},\ \bibinfo {pages} {2 } (\bibinfo {year}
  {1996})}\BibitemShut {NoStop}%
\bibitem [{\citenamefont {Khokonov}(2019)}]{Murat2019}%
  \BibitemOpen
  \bibfield  {author} {\bibinfo {author} {\bibfnamefont {M.}~\bibnamefont
  {Khokonov}},\ }\href {\doibase
  https://doi.org/10.1016/j.physletb.2019.02.034} {\bibfield  {journal}
  {\bibinfo  {journal} {Physics Letters B}\ }\textbf {\bibinfo {volume}
  {791}},\ \bibinfo {pages} {281 } (\bibinfo {year} {2019})}\BibitemShut
  {NoStop}%
\bibitem [{\citenamefont {Khokonov}\ and\ \citenamefont
  {Andersen}(2019)}]{MuratAndersen2019}%
  \BibitemOpen
  \bibfield  {author} {\bibinfo {author} {\bibfnamefont {M.~K.}\ \bibnamefont
  {Khokonov}}\ and\ \bibinfo {author} {\bibfnamefont {J.~U.}\ \bibnamefont
  {Andersen}},\ }\href@noop {} {\bibfield  {journal} {\bibinfo  {journal}
  {Foundations of Physics}\ }\textbf {\bibinfo {volume} {49}},\ \bibinfo
  {pages} {750} (\bibinfo {year} {2019})}\BibitemShut {NoStop}%
\bibitem [{\citenamefont {Vranic}\ \emph {et~al.}(2014)\citenamefont {Vranic},
  \citenamefont {Martins}, \citenamefont {Vieira}, \citenamefont {Fonseca},\
  and\ \citenamefont {Silva}}]{Vranic_2014}%
  \BibitemOpen
  \bibfield  {author} {\bibinfo {author} {\bibfnamefont {M.}~\bibnamefont
  {Vranic}}, \bibinfo {author} {\bibfnamefont {J.~L.}\ \bibnamefont {Martins}},
  \bibinfo {author} {\bibfnamefont {J.}~\bibnamefont {Vieira}}, \bibinfo
  {author} {\bibfnamefont {R.~A.}\ \bibnamefont {Fonseca}}, \ and\ \bibinfo
  {author} {\bibfnamefont {L.~O.}\ \bibnamefont {Silva}},\ }\href {\doibase
  10.1103/PhysRevLett.113.134801} {\bibfield  {journal} {\bibinfo  {journal}
  {Phys. Rev. Lett.}\ }\textbf {\bibinfo {volume} {113}},\ \bibinfo {pages}
  {134801} (\bibinfo {year} {2014})}\BibitemShut {NoStop}%
\bibitem [{\citenamefont {Blackburn}\ \emph {et~al.}(2014)\citenamefont
  {Blackburn}, \citenamefont {Ridgers}, \citenamefont {Kirk},\ and\
  \citenamefont {Bell}}]{Blackburn_2014}%
  \BibitemOpen
  \bibfield  {author} {\bibinfo {author} {\bibfnamefont {T.~G.}\ \bibnamefont
  {Blackburn}}, \bibinfo {author} {\bibfnamefont {C.~P.}\ \bibnamefont
  {Ridgers}}, \bibinfo {author} {\bibfnamefont {J.~G.}\ \bibnamefont {Kirk}}, \
  and\ \bibinfo {author} {\bibfnamefont {A.~R.}\ \bibnamefont {Bell}},\ }\href
  {\doibase 10.1103/PhysRevLett.112.015001} {\bibfield  {journal} {\bibinfo
  {journal} {Phys. Rev. Lett.}\ }\textbf {\bibinfo {volume} {112}},\ \bibinfo
  {pages} {015001} (\bibinfo {year} {2014})}\BibitemShut {NoStop}%
\bibitem [{\citenamefont {Tamburini}\ \emph {et~al.}(2014)\citenamefont
  {Tamburini}, \citenamefont {Keitel},\ and\ \citenamefont
  {Piazza}}]{Tamburini_2014}%
  \BibitemOpen
  \bibfield  {author} {\bibinfo {author} {\bibfnamefont {M.}~\bibnamefont
  {Tamburini}}, \bibinfo {author} {\bibfnamefont {C.~H.}\ \bibnamefont
  {Keitel}}, \ and\ \bibinfo {author} {\bibfnamefont {A.~D.}\ \bibnamefont
  {Piazza}},\ }\href {\doibase 10.1103/PhysRevE.89.021201} {\bibfield
  {journal} {\bibinfo  {journal} {Phys. Rev. E}\ }\textbf {\bibinfo {volume}
  {89}},\ \bibinfo {pages} {021201} (\bibinfo {year} {2014})}\BibitemShut
  {NoStop}%
\bibitem [{\citenamefont {Li}\ \emph {et~al.}(2014)\citenamefont {Li},
  \citenamefont {Hatsagortsyan},\ and\ \citenamefont {Keitel}}]{Li_2014}%
  \BibitemOpen
  \bibfield  {author} {\bibinfo {author} {\bibfnamefont {J.-X.}\ \bibnamefont
  {Li}}, \bibinfo {author} {\bibfnamefont {K.~Z.}\ \bibnamefont
  {Hatsagortsyan}}, \ and\ \bibinfo {author} {\bibfnamefont {C.~H.}\
  \bibnamefont {Keitel}},\ }\href {\doibase 10.1103/PhysRevLett.113.044801}
  {\bibfield  {journal} {\bibinfo  {journal} {Phys. Rev. Lett.}\ }\textbf
  {\bibinfo {volume} {113}},\ \bibinfo {pages} {044801} (\bibinfo {year}
  {2014})}\BibitemShut {NoStop}%
\bibitem [{\citenamefont {Heinzl}\ \emph {et~al.}(2015)\citenamefont {Heinzl},
  \citenamefont {Harvey}, \citenamefont {Ilderton}, \citenamefont {Marklund},
  \citenamefont {Bulanov}, \citenamefont {Rykovanov}, \citenamefont
  {Schroeder}, \citenamefont {Esarey},\ and\ \citenamefont
  {Leemans}}]{Heinzl_2015}%
  \BibitemOpen
  \bibfield  {author} {\bibinfo {author} {\bibfnamefont {T.}~\bibnamefont
  {Heinzl}}, \bibinfo {author} {\bibfnamefont {C.}~\bibnamefont {Harvey}},
  \bibinfo {author} {\bibfnamefont {A.}~\bibnamefont {Ilderton}}, \bibinfo
  {author} {\bibfnamefont {M.}~\bibnamefont {Marklund}}, \bibinfo {author}
  {\bibfnamefont {S.~S.}\ \bibnamefont {Bulanov}}, \bibinfo {author}
  {\bibfnamefont {S.}~\bibnamefont {Rykovanov}}, \bibinfo {author}
  {\bibfnamefont {C.~B.}\ \bibnamefont {Schroeder}}, \bibinfo {author}
  {\bibfnamefont {E.}~\bibnamefont {Esarey}}, \ and\ \bibinfo {author}
  {\bibfnamefont {W.~P.}\ \bibnamefont {Leemans}},\ }\href {\doibase
  10.1103/PhysRevE.91.023207} {\bibfield  {journal} {\bibinfo  {journal} {Phys.
  Rev. E}\ }\textbf {\bibinfo {volume} {91}},\ \bibinfo {pages} {023207}
  (\bibinfo {year} {2015})}\BibitemShut {NoStop}%
\bibitem [{\citenamefont {Yoffe}\ \emph {et~al.}(2015)\citenamefont {Yoffe},
  \citenamefont {Kravets}, \citenamefont {Noble},\ and\ \citenamefont
  {Jaroszynski}}]{Yoffe_2015}%
  \BibitemOpen
  \bibfield  {author} {\bibinfo {author} {\bibfnamefont {S.~R.}\ \bibnamefont
  {Yoffe}}, \bibinfo {author} {\bibfnamefont {Y.}~\bibnamefont {Kravets}},
  \bibinfo {author} {\bibfnamefont {A.}~\bibnamefont {Noble}}, \ and\ \bibinfo
  {author} {\bibfnamefont {D.~A.}\ \bibnamefont {Jaroszynski}},\ }\href
  {\doibase 10.1088/1367-2630/17/5/053025} {\bibfield  {journal} {\bibinfo
  {journal} {New Journal of Physics}\ }\textbf {\bibinfo {volume} {17}},\
  \bibinfo {pages} {053025} (\bibinfo {year} {2015})}\BibitemShut {NoStop}%
\bibitem [{\citenamefont {Capdessus}\ and\ \citenamefont
  {McKenna}(2015)}]{Capdessus_2015}%
  \BibitemOpen
  \bibfield  {author} {\bibinfo {author} {\bibfnamefont {R.}~\bibnamefont
  {Capdessus}}\ and\ \bibinfo {author} {\bibfnamefont {P.}~\bibnamefont
  {McKenna}},\ }\href {\doibase 10.1103/PhysRevE.91.053105} {\bibfield
  {journal} {\bibinfo  {journal} {Phys. Rev. E}\ }\textbf {\bibinfo {volume}
  {91}},\ \bibinfo {pages} {053105} (\bibinfo {year} {2015})}\BibitemShut
  {NoStop}%
\bibitem [{\citenamefont {Vranic}\ \emph {et~al.}(2016)\citenamefont {Vranic},
  \citenamefont {Grismayer}, \citenamefont {Fonseca},\ and\ \citenamefont
  {Silva}}]{Vranic_2016}%
  \BibitemOpen
  \bibfield  {author} {\bibinfo {author} {\bibfnamefont {M.}~\bibnamefont
  {Vranic}}, \bibinfo {author} {\bibfnamefont {T.}~\bibnamefont {Grismayer}},
  \bibinfo {author} {\bibfnamefont {R.~A.}\ \bibnamefont {Fonseca}}, \ and\
  \bibinfo {author} {\bibfnamefont {L.~O.}\ \bibnamefont {Silva}},\ }\href
  {\doibase 10.1088/1367-2630/18/7/073035} {\bibfield  {journal} {\bibinfo
  {journal} {New Journal of Physics}\ }\textbf {\bibinfo {volume} {18}},\
  \bibinfo {pages} {073035} (\bibinfo {year} {2016})}\BibitemShut {NoStop}%
\bibitem [{\citenamefont {Dinu}\ \emph {et~al.}(2016)\citenamefont {Dinu},
  \citenamefont {Harvey}, \citenamefont {Ilderton}, \citenamefont {Marklund},\
  and\ \citenamefont {Torgrimsson}}]{Dinu_2016}%
  \BibitemOpen
  \bibfield  {author} {\bibinfo {author} {\bibfnamefont {V.}~\bibnamefont
  {Dinu}}, \bibinfo {author} {\bibfnamefont {C.}~\bibnamefont {Harvey}},
  \bibinfo {author} {\bibfnamefont {A.}~\bibnamefont {Ilderton}}, \bibinfo
  {author} {\bibfnamefont {M.}~\bibnamefont {Marklund}}, \ and\ \bibinfo
  {author} {\bibfnamefont {G.}~\bibnamefont {Torgrimsson}},\ }\href {\doibase
  10.1103/PhysRevLett.116.044801} {\bibfield  {journal} {\bibinfo  {journal}
  {Phys. Rev. Lett.}\ }\textbf {\bibinfo {volume} {116}},\ \bibinfo {pages}
  {044801} (\bibinfo {year} {2016})}\BibitemShut {NoStop}%
\bibitem [{\citenamefont {Piazza}\ \emph {et~al.}(2017)\citenamefont {Piazza},
  \citenamefont {Wistisen},\ and\ \citenamefont
  {Uggerh{\o}j}}]{Di_Piazza_2017}%
  \BibitemOpen
  \bibfield  {author} {\bibinfo {author} {\bibfnamefont {A.~D.}\ \bibnamefont
  {Piazza}}, \bibinfo {author} {\bibfnamefont {T.~N.}\ \bibnamefont
  {Wistisen}}, \ and\ \bibinfo {author} {\bibfnamefont {U.~I.}\ \bibnamefont
  {Uggerh{\o}j}},\ }\href {\doibase
  https://doi.org/10.1016/j.physletb.2016.10.083} {\bibfield  {journal}
  {\bibinfo  {journal} {Physics Letters B}\ }\textbf {\bibinfo {volume}
  {765}},\ \bibinfo {pages} {1 } (\bibinfo {year} {2017})}\BibitemShut
  {NoStop}%
\bibitem [{\citenamefont {Harvey}\ \emph {et~al.}(2017)\citenamefont {Harvey},
  \citenamefont {Gonoskov}, \citenamefont {Ilderton},\ and\ \citenamefont
  {Marklund}}]{Harvey_2017}%
  \BibitemOpen
  \bibfield  {author} {\bibinfo {author} {\bibfnamefont {C.~N.}\ \bibnamefont
  {Harvey}}, \bibinfo {author} {\bibfnamefont {A.}~\bibnamefont {Gonoskov}},
  \bibinfo {author} {\bibfnamefont {A.}~\bibnamefont {Ilderton}}, \ and\
  \bibinfo {author} {\bibfnamefont {M.}~\bibnamefont {Marklund}},\ }\href
  {\doibase 10.1103/PhysRevLett.118.105004} {\bibfield  {journal} {\bibinfo
  {journal} {Phys. Rev. Lett.}\ }\textbf {\bibinfo {volume} {118}},\ \bibinfo
  {pages} {105004} (\bibinfo {year} {2017})}\BibitemShut {NoStop}%
\bibitem [{\citenamefont {Ridgers}\ \emph {et~al.}(2017)\citenamefont
  {Ridgers}, \citenamefont {Blackburn}, \citenamefont {Del~Sorbo},
  \citenamefont {Bradley}, \citenamefont {Slade-Lowther}, \citenamefont
  {Baird}, \citenamefont {Mangles}, \citenamefont {McKenna}, \citenamefont
  {Marklund}, \citenamefont {Murphy},\ and\ \citenamefont
  {et~al.}}]{Ridgers_2017}%
  \BibitemOpen
  \bibfield  {author} {\bibinfo {author} {\bibfnamefont {C.~P.}\ \bibnamefont
  {Ridgers}}, \bibinfo {author} {\bibfnamefont {T.~G.}\ \bibnamefont
  {Blackburn}}, \bibinfo {author} {\bibfnamefont {D.}~\bibnamefont
  {Del~Sorbo}}, \bibinfo {author} {\bibfnamefont {L.~E.}\ \bibnamefont
  {Bradley}}, \bibinfo {author} {\bibfnamefont {C.}~\bibnamefont
  {Slade-Lowther}}, \bibinfo {author} {\bibfnamefont {C.~D.}\ \bibnamefont
  {Baird}}, \bibinfo {author} {\bibfnamefont {S.~P.~D.}\ \bibnamefont
  {Mangles}}, \bibinfo {author} {\bibfnamefont {P.}~\bibnamefont {McKenna}},
  \bibinfo {author} {\bibfnamefont {M.}~\bibnamefont {Marklund}}, \bibinfo
  {author} {\bibfnamefont {C.~D.}\ \bibnamefont {Murphy}}, \ and\ \bibinfo
  {author} {\bibnamefont {et~al.}},\ }\href {\doibase
  10.1017/S0022377817000642} {\bibfield  {journal} {\bibinfo  {journal}
  {Journal of Plasma Physics}\ }\textbf {\bibinfo {volume} {83}},\ \bibinfo
  {pages} {715830502} (\bibinfo {year} {2017})}\BibitemShut {NoStop}%
\bibitem [{\citenamefont {Niel}\ \emph
  {et~al.}(2018{\natexlab{a}})\citenamefont {Niel}, \citenamefont {Riconda},
  \citenamefont {Amiranoff}, \citenamefont {Duclous},\ and\ \citenamefont
  {Grech}}]{Niel_2017}%
  \BibitemOpen
  \bibfield  {author} {\bibinfo {author} {\bibfnamefont {F.}~\bibnamefont
  {Niel}}, \bibinfo {author} {\bibfnamefont {C.}~\bibnamefont {Riconda}},
  \bibinfo {author} {\bibfnamefont {F.}~\bibnamefont {Amiranoff}}, \bibinfo
  {author} {\bibfnamefont {R.}~\bibnamefont {Duclous}}, \ and\ \bibinfo
  {author} {\bibfnamefont {M.}~\bibnamefont {Grech}},\ }\href {\doibase
  10.1103/PhysRevE.97.043209} {\bibfield  {journal} {\bibinfo  {journal} {Phys.
  Rev. E}\ }\textbf {\bibinfo {volume} {97}},\ \bibinfo {pages} {043209}
  (\bibinfo {year} {2018}{\natexlab{a}})}\BibitemShut {NoStop}%
\bibitem [{\citenamefont {Niel}\ \emph
  {et~al.}(2018{\natexlab{b}})\citenamefont {Niel}, \citenamefont {Riconda},
  \citenamefont {Amiranoff}, \citenamefont {Lobet}, \citenamefont {Derouillat},
  \citenamefont {Pérez}, \citenamefont {Vinci},\ and\ \citenamefont
  {Grech}}]{Niel_2018}%
  \BibitemOpen
  \bibfield  {author} {\bibinfo {author} {\bibfnamefont {F.}~\bibnamefont
  {Niel}}, \bibinfo {author} {\bibfnamefont {C.}~\bibnamefont {Riconda}},
  \bibinfo {author} {\bibfnamefont {F.}~\bibnamefont {Amiranoff}}, \bibinfo
  {author} {\bibfnamefont {M.}~\bibnamefont {Lobet}}, \bibinfo {author}
  {\bibfnamefont {F.}~\bibnamefont {Derouillat}}, \bibinfo {author}
  {\bibfnamefont {T.}~\bibnamefont {Pérez}}, \bibinfo {author} {\bibfnamefont
  {T.}~\bibnamefont {Vinci}}, \ and\ \bibinfo {author} {\bibfnamefont
  {M.}~\bibnamefont {Grech}},\ }\href {\doibase 10.1088/1361-6587/aace22}
  {\bibfield  {journal} {\bibinfo  {journal} {arXiv:1802.02927
  [physics.plasm-ph]}\ } (\bibinfo {year} {2018}{\natexlab{b}}),\
  10.1088/1361-6587/aace22}\BibitemShut {NoStop}%
\bibitem [{\citenamefont {Wistisen}\ \emph {et~al.}(2018)\citenamefont
  {Wistisen}, \citenamefont {Di~Piazza}, \citenamefont {Knudsen},\ and\
  \citenamefont {Uggerh{\o}j}}]{Wistisen_2018}%
  \BibitemOpen
  \bibfield  {author} {\bibinfo {author} {\bibfnamefont {T.~N.}\ \bibnamefont
  {Wistisen}}, \bibinfo {author} {\bibfnamefont {A.}~\bibnamefont {Di~Piazza}},
  \bibinfo {author} {\bibfnamefont {H.~V.}\ \bibnamefont {Knudsen}}, \ and\
  \bibinfo {author} {\bibfnamefont {U.~I.}\ \bibnamefont {Uggerh{\o}j}},\
  }\href {\doibase 10.1038/s41467-018-03165-4} {\bibfield  {journal} {\bibinfo
  {journal} {Nature Communications}\ }\textbf {\bibinfo {volume} {9}},\
  \bibinfo {pages} {795} (\bibinfo {year} {2018})}\BibitemShut {NoStop}%
\bibitem [{\citenamefont {Cole}\ \emph {et~al.}(2018)\citenamefont {Cole},
  \citenamefont {Behm}, \citenamefont {Gerstmayr}, \citenamefont {Blackburn},
  \citenamefont {Wood}, \citenamefont {Baird}, \citenamefont {Duff},
  \citenamefont {Harvey}, \citenamefont {Ilderton}, \citenamefont {Joglekar},
  \citenamefont {Krushelnick}, \citenamefont {Kuschel}, \citenamefont
  {Marklund}, \citenamefont {McKenna}, \citenamefont {Murphy}, \citenamefont
  {Poder}, \citenamefont {Ridgers}, \citenamefont {Samarin}, \citenamefont
  {Sarri}, \citenamefont {Symes}, \citenamefont {Thomas}, \citenamefont
  {Warwick}, \citenamefont {Zepf}, \citenamefont {Najmudin},\ and\
  \citenamefont {Mangles}}]{Cole_2018}%
  \BibitemOpen
  \bibfield  {author} {\bibinfo {author} {\bibfnamefont {J.~M.}\ \bibnamefont
  {Cole}}, \bibinfo {author} {\bibfnamefont {K.~T.}\ \bibnamefont {Behm}},
  \bibinfo {author} {\bibfnamefont {E.}~\bibnamefont {Gerstmayr}}, \bibinfo
  {author} {\bibfnamefont {T.~G.}\ \bibnamefont {Blackburn}}, \bibinfo {author}
  {\bibfnamefont {J.~C.}\ \bibnamefont {Wood}}, \bibinfo {author}
  {\bibfnamefont {C.~D.}\ \bibnamefont {Baird}}, \bibinfo {author}
  {\bibfnamefont {M.~J.}\ \bibnamefont {Duff}}, \bibinfo {author}
  {\bibfnamefont {C.}~\bibnamefont {Harvey}}, \bibinfo {author} {\bibfnamefont
  {A.}~\bibnamefont {Ilderton}}, \bibinfo {author} {\bibfnamefont {A.~S.}\
  \bibnamefont {Joglekar}}, \bibinfo {author} {\bibfnamefont {K.}~\bibnamefont
  {Krushelnick}}, \bibinfo {author} {\bibfnamefont {S.}~\bibnamefont
  {Kuschel}}, \bibinfo {author} {\bibfnamefont {M.}~\bibnamefont {Marklund}},
  \bibinfo {author} {\bibfnamefont {P.}~\bibnamefont {McKenna}}, \bibinfo
  {author} {\bibfnamefont {C.~D.}\ \bibnamefont {Murphy}}, \bibinfo {author}
  {\bibfnamefont {K.}~\bibnamefont {Poder}}, \bibinfo {author} {\bibfnamefont
  {C.~P.}\ \bibnamefont {Ridgers}}, \bibinfo {author} {\bibfnamefont {G.~M.}\
  \bibnamefont {Samarin}}, \bibinfo {author} {\bibfnamefont {G.}~\bibnamefont
  {Sarri}}, \bibinfo {author} {\bibfnamefont {D.~R.}\ \bibnamefont {Symes}},
  \bibinfo {author} {\bibfnamefont {A.~G.~R.}\ \bibnamefont {Thomas}}, \bibinfo
  {author} {\bibfnamefont {J.}~\bibnamefont {Warwick}}, \bibinfo {author}
  {\bibfnamefont {M.}~\bibnamefont {Zepf}}, \bibinfo {author} {\bibfnamefont
  {Z.}~\bibnamefont {Najmudin}}, \ and\ \bibinfo {author} {\bibfnamefont
  {S.~P.~D.}\ \bibnamefont {Mangles}},\ }\href {\doibase
  10.1103/PhysRevX.8.011020} {\bibfield  {journal} {\bibinfo  {journal} {Phys.
  Rev. X}\ }\textbf {\bibinfo {volume} {8}},\ \bibinfo {pages} {011020}
  (\bibinfo {year} {2018})}\BibitemShut {NoStop}%
\bibitem [{\citenamefont {Poder}\ \emph {et~al.}(2018)\citenamefont {Poder},
  \citenamefont {Tamburini}, \citenamefont {Sarri}, \citenamefont {Di~Piazza},
  \citenamefont {Kuschel}, \citenamefont {Baird}, \citenamefont {Behm},
  \citenamefont {Bohlen}, \citenamefont {Cole}, \citenamefont {Corvan},
  \citenamefont {Duff}, \citenamefont {Gerstmayr}, \citenamefont {Keitel},
  \citenamefont {Krushelnick}, \citenamefont {Mangles}, \citenamefont
  {McKenna}, \citenamefont {Murphy}, \citenamefont {Najmudin}, \citenamefont
  {Ridgers}, \citenamefont {Samarin}, \citenamefont {Symes}, \citenamefont
  {Thomas}, \citenamefont {Warwick},\ and\ \citenamefont {Zepf}}]{Poder_2017}%
  \BibitemOpen
  \bibfield  {author} {\bibinfo {author} {\bibfnamefont {K.}~\bibnamefont
  {Poder}}, \bibinfo {author} {\bibfnamefont {M.}~\bibnamefont {Tamburini}},
  \bibinfo {author} {\bibfnamefont {G.}~\bibnamefont {Sarri}}, \bibinfo
  {author} {\bibfnamefont {A.}~\bibnamefont {Di~Piazza}}, \bibinfo {author}
  {\bibfnamefont {S.}~\bibnamefont {Kuschel}}, \bibinfo {author} {\bibfnamefont
  {C.~D.}\ \bibnamefont {Baird}}, \bibinfo {author} {\bibfnamefont
  {K.}~\bibnamefont {Behm}}, \bibinfo {author} {\bibfnamefont {S.}~\bibnamefont
  {Bohlen}}, \bibinfo {author} {\bibfnamefont {J.~M.}\ \bibnamefont {Cole}},
  \bibinfo {author} {\bibfnamefont {D.~J.}\ \bibnamefont {Corvan}}, \bibinfo
  {author} {\bibfnamefont {M.}~\bibnamefont {Duff}}, \bibinfo {author}
  {\bibfnamefont {E.}~\bibnamefont {Gerstmayr}}, \bibinfo {author}
  {\bibfnamefont {C.~H.}\ \bibnamefont {Keitel}}, \bibinfo {author}
  {\bibfnamefont {K.}~\bibnamefont {Krushelnick}}, \bibinfo {author}
  {\bibfnamefont {S.~P.~D.}\ \bibnamefont {Mangles}}, \bibinfo {author}
  {\bibfnamefont {P.}~\bibnamefont {McKenna}}, \bibinfo {author} {\bibfnamefont
  {C.~D.}\ \bibnamefont {Murphy}}, \bibinfo {author} {\bibfnamefont
  {Z.}~\bibnamefont {Najmudin}}, \bibinfo {author} {\bibfnamefont {C.~P.}\
  \bibnamefont {Ridgers}}, \bibinfo {author} {\bibfnamefont {G.~M.}\
  \bibnamefont {Samarin}}, \bibinfo {author} {\bibfnamefont {D.~R.}\
  \bibnamefont {Symes}}, \bibinfo {author} {\bibfnamefont {A.~G.~R.}\
  \bibnamefont {Thomas}}, \bibinfo {author} {\bibfnamefont {J.}~\bibnamefont
  {Warwick}}, \ and\ \bibinfo {author} {\bibfnamefont {M.}~\bibnamefont
  {Zepf}},\ }\href {\doibase 10.1103/PhysRevX.8.031004} {\bibfield  {journal}
  {\bibinfo  {journal} {Phys. Rev. X}\ }\textbf {\bibinfo {volume} {8}},\
  \bibinfo {pages} {031004} (\bibinfo {year} {2018})}\BibitemShut {NoStop}%
\bibitem [{\citenamefont {Hammond}(2010)}]{Hammond_2010}%
  \BibitemOpen
  \bibfield  {author} {\bibinfo {author} {\bibfnamefont {R.}~\bibnamefont
  {Hammond}},\ }\href@noop {} {\bibfield  {journal} {\bibinfo  {journal}
  {Electronic Journal of Theoretical Physics}\ }\textbf {\bibinfo {volume}
  {6}},\ \bibinfo {pages} {221} (\bibinfo {year} {2010})}\BibitemShut {NoStop}%
\bibitem [{\citenamefont {Di~Piazza}\ \emph {et~al.}(2012)\citenamefont
  {Di~Piazza}, \citenamefont {M\"uller}, \citenamefont {Hatsagortsyan},\ and\
  \citenamefont {Keitel}}]{Di_Piazza_2012}%
  \BibitemOpen
  \bibfield  {author} {\bibinfo {author} {\bibfnamefont {A.}~\bibnamefont
  {Di~Piazza}}, \bibinfo {author} {\bibfnamefont {C.}~\bibnamefont {M\"uller}},
  \bibinfo {author} {\bibfnamefont {K.~Z.}\ \bibnamefont {Hatsagortsyan}}, \
  and\ \bibinfo {author} {\bibfnamefont {C.~H.}\ \bibnamefont {Keitel}},\
  }\href {\doibase 10.1103/RevModPhys.84.1177} {\bibfield  {journal} {\bibinfo
  {journal} {Rev. Mod. Phys.}\ }\textbf {\bibinfo {volume} {84}},\ \bibinfo
  {pages} {1177} (\bibinfo {year} {2012})}\BibitemShut {NoStop}%
\bibitem [{\citenamefont {Burton}\ and\ \citenamefont
  {Noble}(2014)}]{Burton_2014}%
  \BibitemOpen
  \bibfield  {author} {\bibinfo {author} {\bibfnamefont {D.}~\bibnamefont
  {Burton}}\ and\ \bibinfo {author} {\bibfnamefont {A.}~\bibnamefont {Noble}},\
  }\href {\doibase 10.1080/00107514.2014.886840} {\bibfield  {journal}
  {\bibinfo  {journal} {Contemporary Physics}\ }\textbf {\bibinfo {volume}
  {55}},\ \bibinfo {pages} {110} (\bibinfo {year} {2014})}\BibitemShut
  {NoStop}%
\bibitem [{\citenamefont {Wistisen}\ \emph {et~al.}(2019)\citenamefont
  {Wistisen}, \citenamefont {Di~Piazza}, \citenamefont {Nielsen}, \citenamefont
  {S\o{}rensen},\ and\ \citenamefont {Uggerh\o{}j}}]{cern2017}%
  \BibitemOpen
  \bibfield  {author} {\bibinfo {author} {\bibfnamefont {T.~N.}\ \bibnamefont
  {Wistisen}}, \bibinfo {author} {\bibfnamefont {A.}~\bibnamefont {Di~Piazza}},
  \bibinfo {author} {\bibfnamefont {C.~F.}\ \bibnamefont {Nielsen}}, \bibinfo
  {author} {\bibfnamefont {A.~H.}\ \bibnamefont {S\o{}rensen}}, \ and\ \bibinfo
  {author} {\bibfnamefont {U.~I.}\ \bibnamefont {Uggerh\o{}j}} (\bibinfo
  {collaboration} {CERN NA63}),\ }\href {\doibase
  10.1103/PhysRevResearch.1.033014} {\bibfield  {journal} {\bibinfo  {journal}
  {Phys. Rev. Research}\ }\textbf {\bibinfo {volume} {1}},\ \bibinfo {pages}
  {033014} (\bibinfo {year} {2019})}\BibitemShut {NoStop}%
\bibitem [{\citenamefont {Berestetskii}\ \emph {et~al.}(1989)\citenamefont
  {Berestetskii}, \citenamefont {Lifshitz},\ and\ \citenamefont
  {Pitaevskii}}]{Berestetskii_b_1989}%
  \BibitemOpen
  \bibfield  {author} {\bibinfo {author} {\bibfnamefont {V.~B.}\ \bibnamefont
  {Berestetskii}}, \bibinfo {author} {\bibfnamefont {E.~M.}\ \bibnamefont
  {Lifshitz}}, \ and\ \bibinfo {author} {\bibfnamefont {L.~P.}\ \bibnamefont
  {Pitaevskii}},\ }\href@noop {} {\emph {\bibinfo {title} {Quantum
  Electrodynamics}}}\ (\bibinfo  {publisher} {Pergamon, New York},\ \bibinfo
  {year} {1989})\BibitemShut {NoStop}%
\bibitem [{\citenamefont {Khokonov}(1992)}]{Murat1992}%
  \BibitemOpen
  \bibfield  {author} {\bibinfo {author} {\bibfnamefont {M.}~\bibnamefont
  {Khokonov}},\ }\href@noop {} {\bibfield  {journal} {\bibinfo  {journal} {JETP
  Letters}\ }\textbf {\bibinfo {volume} {56}},\ \bibinfo {pages} {333}
  (\bibinfo {year} {1992})}\BibitemShut {NoStop}%
\bibitem [{\citenamefont {Khokonov}(1993)}]{Murat1993}%
  \BibitemOpen
  \bibfield  {author} {\bibinfo {author} {\bibfnamefont {M.}~\bibnamefont
  {Khokonov}},\ }\href@noop {} {\bibfield  {journal} {\bibinfo  {journal}
  {Journal of Experimental and Theoretical Physics - JETP}\ }\textbf {\bibinfo
  {volume} {76}},\ \bibinfo {pages} {849} (\bibinfo {year} {1993})}\BibitemShut
  {NoStop}%
\bibitem [{\citenamefont {Lindhard}(1965)}]{Lind65}%
  \BibitemOpen
  \bibfield  {author} {\bibinfo {author} {\bibfnamefont {J.}~\bibnamefont
  {Lindhard}},\ }\href@noop {} {\bibfield  {journal} {\bibinfo  {journal} {Mat.
  Fys. Medd. Dan. Vid. Selsk.}\ }\textbf {\bibinfo {volume} {34}},\ \bibinfo
  {pages} {no. 14, 1} (\bibinfo {year} {1965})}\BibitemShut {NoStop}%
\bibitem [{\citenamefont {Andersen}(2018)}]{JUAnotes}%
  \BibitemOpen
  \bibfield  {author} {\bibinfo {author} {\bibfnamefont {J.~U.}\ \bibnamefont
  {Andersen}},\ }\href
  {https://phys.au.dk/forskning/publikationer/lecture-notes/} {\enquote
  {\bibinfo {title} {{Notes on channeling}},}\ } (\bibinfo {year} {2018}),\
  \bibinfo {note} {lecture notes, Aarhus University,
  https://phys.au.dk/forskning/publikationer/lecture-notes/}\BibitemShut
  {NoStop}%
\bibitem [{\citenamefont {Doyle}\ and\ \citenamefont {Turner}(1968)}]{Doyl68}%
  \BibitemOpen
  \bibfield  {author} {\bibinfo {author} {\bibfnamefont {P.~A.}\ \bibnamefont
  {Doyle}}\ and\ \bibinfo {author} {\bibfnamefont {P.~S.}\ \bibnamefont
  {Turner}},\ }\href {https://doi.org/10.1107/S0567739468000756} {\bibfield
  {journal} {\bibinfo  {journal} {Acta Crystallographica Section A}\ }\textbf
  {\bibinfo {volume} {24}},\ \bibinfo {pages} {390} (\bibinfo {year}
  {1968})}\BibitemShut {NoStop}%
\bibitem [{\citenamefont {Andersen}\ \emph {et~al.}(1982)\citenamefont
  {Andersen}, \citenamefont {Bonderup}, \citenamefont {L{\ae}gsgaard},
  \citenamefont {Marsh},\ and\ \citenamefont {S{\o}rensen}}]{Ande82}%
  \BibitemOpen
  \bibfield  {author} {\bibinfo {author} {\bibfnamefont {J.~U.}\ \bibnamefont
  {Andersen}}, \bibinfo {author} {\bibfnamefont {E.}~\bibnamefont {Bonderup}},
  \bibinfo {author} {\bibfnamefont {E.}~\bibnamefont {L{\ae}gsgaard}}, \bibinfo
  {author} {\bibfnamefont {B.~B.}\ \bibnamefont {Marsh}}, \ and\ \bibinfo
  {author} {\bibfnamefont {A.~H.}\ \bibnamefont {S{\o}rensen}},\ }\href
  {\doibase https://doi.org/10.1016/0029-554X(82)90517-1} {\bibfield  {journal}
  {\bibinfo  {journal} {Nuclear Instruments and Methods in Physics Research}\
  }\textbf {\bibinfo {volume} {194}},\ \bibinfo {pages} {209 } (\bibinfo {year}
  {1982})}\BibitemShut {NoStop}%
\bibitem [{\citenamefont {Nielsen}\ and\ \citenamefont
  {Weber}(1980)}]{Nielsen_1980}%
  \BibitemOpen
  \bibfield  {author} {\bibinfo {author} {\bibfnamefont {O.~H.}\ \bibnamefont
  {Nielsen}}\ and\ \bibinfo {author} {\bibfnamefont {W.}~\bibnamefont
  {Weber}},\ }\href {\doibase 10.1088/0022-3719/13/13/005} {\bibfield
  {journal} {\bibinfo  {journal} {Journal of Physics C: Solid State Physics}\
  }\textbf {\bibinfo {volume} {13}},\ \bibinfo {pages} {2449} (\bibinfo {year}
  {1980})}\BibitemShut {NoStop}%
\bibitem [{\citenamefont {Korol}\ \emph {et~al.}(2013)\citenamefont {Korol},
  \citenamefont {Solov'yov},\ and\ \citenamefont {Greiner}}]{Koro13}%
  \BibitemOpen
  \bibfield  {author} {\bibinfo {author} {\bibfnamefont {A.~V.}\ \bibnamefont
  {Korol}}, \bibinfo {author} {\bibfnamefont {A.~V.}\ \bibnamefont
  {Solov'yov}}, \ and\ \bibinfo {author} {\bibfnamefont {W.}~\bibnamefont
  {Greiner}},\ }\href@noop {} {\emph {\bibinfo {title} {Channeling and
  Radiation in Periodically Bent Crystals, Springer Series on Atomic, Optical
  and Plasma Physics 69}}}\ (\bibinfo  {publisher} {Springer-Verlag Berlin
  Heidelberg},\ \bibinfo {year} {2013})\BibitemShut {NoStop}%
\bibitem [{\citenamefont {Tanabashi~{\em{et al.\/}}}(2018)}]{PDG_2018}%
  \BibitemOpen
  \bibfield  {author} {\bibinfo {author} {\bibfnamefont {M.}~\bibnamefont
  {Tanabashi~{\em{et al.\/}}}} (\bibinfo {collaboration} {Particle Data
  Group}),\ }\href {\doibase 10.1103/PhysRevD.98.030001} {\bibfield  {journal}
  {\bibinfo  {journal} {Phys. Rev. D}\ }\textbf {\bibinfo {volume} {98}},\
  \bibinfo {pages} {030001} (\bibinfo {year} {2018})}\BibitemShut {NoStop}%
\bibitem [{\citenamefont {Lynch}\ and\ \citenamefont
  {Dahl}(1991)}]{LynchDahl1991}%
  \BibitemOpen
  \bibfield  {author} {\bibinfo {author} {\bibfnamefont {G.~R.}\ \bibnamefont
  {Lynch}}\ and\ \bibinfo {author} {\bibfnamefont {O.~I.}\ \bibnamefont
  {Dahl}},\ }\href@noop {} {\bibfield  {journal} {\bibinfo  {journal} {Nuclear
  Instruments and Methods in Physics Research B}\ }\textbf {\bibinfo {volume}
  {58}},\ \bibinfo {pages} {6 } (\bibinfo {year} {1991})}\BibitemShut {NoStop}%
\bibitem [{\citenamefont {Taratin}(1998)}]{Taratin1998}%
  \BibitemOpen
  \bibfield  {author} {\bibinfo {author} {\bibfnamefont {A.~M.}\ \bibnamefont
  {Taratin}},\ }\href@noop {} {\bibfield  {journal} {\bibinfo  {journal} {Phys.
  Part. Nucl}\ }\textbf {\bibinfo {volume} {29}},\ \bibinfo {pages} {437}
  (\bibinfo {year} {1998})}\BibitemShut {NoStop}%
\bibitem [{\citenamefont {Bonderup}(1981)}]{EBblue}%
  \BibitemOpen
  \bibfield  {author} {\bibinfo {author} {\bibfnamefont {E.}~\bibnamefont
  {Bonderup}},\ }\href
  {https://phys.au.dk/forskning/publikationer/lecture-notes/} {\enquote
  {\bibinfo {title} {{Penetration of charged particles through matter}},}\ }
  (\bibinfo {year} {1981}),\ \bibinfo {note} {lecture notes, Aarhus University,
  https://phys.au.dk/forskning/publikationer/lecture-notes/}\BibitemShut
  {NoStop}%
\bibitem [{\citenamefont {Jensen}\ and\ \citenamefont
  {S\o{}rensen}(2013)}]{Jens2013}%
  \BibitemOpen
  \bibfield  {author} {\bibinfo {author} {\bibfnamefont {T.~V.}\ \bibnamefont
  {Jensen}}\ and\ \bibinfo {author} {\bibfnamefont {A.~H.}\ \bibnamefont
  {S\o{}rensen}},\ }\href {\doibase 10.1103/PhysRevA.87.022902} {\bibfield
  {journal} {\bibinfo  {journal} {Phys. Rev. A}\ }\textbf {\bibinfo {volume}
  {87}},\ \bibinfo {pages} {022902} (\bibinfo {year} {2013})}\BibitemShut
  {NoStop}%
\bibitem [{\citenamefont {Lindhard}(1991)}]{Lind91}%
  \BibitemOpen
  \bibfield  {author} {\bibinfo {author} {\bibfnamefont {J.}~\bibnamefont
  {Lindhard}},\ }\href {\doibase 10.1103/PhysRevA.43.6032} {\bibfield
  {journal} {\bibinfo  {journal} {Phys. Rev. A}\ }\textbf {\bibinfo {volume}
  {43}},\ \bibinfo {pages} {6032} (\bibinfo {year} {1991})}\BibitemShut
  {NoStop}%
\bibitem [{\citenamefont {Williams}(1935)}]{WeizsackerWilliams1934}%
  \BibitemOpen
  \bibfield  {author} {\bibinfo {author} {\bibfnamefont {E.}~\bibnamefont
  {Williams}},\ }\href@noop {} {\bibfield  {journal} {\bibinfo  {journal}
  {{Mat. Fys. Medd. Dan. Vid. Selsk.}}\ }\textbf {\bibinfo {volume} {13}},\
  \bibinfo {pages} {no. 4, 1} (\bibinfo {year} {1935})}\BibitemShut {NoStop}%
\bibitem [{\citenamefont {Matveev}(1957)}]{Matveev1957}%
  \BibitemOpen
  \bibfield  {author} {\bibinfo {author} {\bibfnamefont {A.~N.}\ \bibnamefont
  {Matveev}},\ }\href@noop {} {\bibfield  {journal} {\bibinfo  {journal}
  {JETP}\ }\textbf {\bibinfo {volume} {4}},\ \bibinfo {pages} {409} (\bibinfo
  {year} {1957})}\BibitemShut {NoStop}%
\bibitem [{\citenamefont {Belkacem}\ \emph {et~al.}(1985)\citenamefont
  {Belkacem}, \citenamefont {Cue},\ and\ \citenamefont
  {Kimball}}]{belkacem_1985}%
  \BibitemOpen
  \bibfield  {author} {\bibinfo {author} {\bibfnamefont {A.}~\bibnamefont
  {Belkacem}}, \bibinfo {author} {\bibfnamefont {N.}~\bibnamefont {Cue}}, \
  and\ \bibinfo {author} {\bibfnamefont {J.}~\bibnamefont {Kimball}},\ }\href
  {\doibase https://doi.org/10.1016/0375-9601(85)90811-4} {\bibfield  {journal}
  {\bibinfo  {journal} {Physics Letters A}\ }\textbf {\bibinfo {volume}
  {111}},\ \bibinfo {pages} {86 } (\bibinfo {year} {1985})}\BibitemShut
  {NoStop}%
\bibitem [{\citenamefont {Kimball}\ \emph {et~al.}(1986)\citenamefont
  {Kimball}, \citenamefont {Cue},\ and\ \citenamefont
  {Belkacem}}]{kimball_1986}%
  \BibitemOpen
  \bibfield  {author} {\bibinfo {author} {\bibfnamefont {J.}~\bibnamefont
  {Kimball}}, \bibinfo {author} {\bibfnamefont {N.}~\bibnamefont {Cue}}, \ and\
  \bibinfo {author} {\bibfnamefont {A.}~\bibnamefont {Belkacem}},\ }\href
  {\doibase https://doi.org/10.1016/0168-583X(86)90461-1} {\bibfield  {journal}
  {\bibinfo  {journal} {Nuclear Instruments and Methods in Physics Research
  Section B: Beam Interactions with Materials and Atoms}\ }\textbf {\bibinfo
  {volume} {13}},\ \bibinfo {pages} {1 } (\bibinfo {year} {1986})}\BibitemShut
  {NoStop}%
\bibitem [{\citenamefont {Wistisen}(2015)}]{Tobias_2015}%
  \BibitemOpen
  \bibfield  {author} {\bibinfo {author} {\bibfnamefont {T.~N.}\ \bibnamefont
  {Wistisen}},\ }\href {\doibase 10.1103/PhysRevD.92.045045} {\bibfield
  {journal} {\bibinfo  {journal} {Phys. Rev. D}\ }\textbf {\bibinfo {volume}
  {92}},\ \bibinfo {pages} {045045} (\bibinfo {year} {2015})}\BibitemShut
  {NoStop}%
\bibitem [{\citenamefont {Bell}(1958)}]{Bell1958}%
  \BibitemOpen
  \bibfield  {author} {\bibinfo {author} {\bibfnamefont {J.}~\bibnamefont
  {Bell}},\ }\href {\doibase https://doi.org/10.1016/0029-5582(58)90185-8}
  {\bibfield  {journal} {\bibinfo  {journal} {Nuclear Physics}\ }\textbf
  {\bibinfo {volume} {8}},\ \bibinfo {pages} {613 } (\bibinfo {year}
  {1958})}\BibitemShut {NoStop}%
\bibitem [{\citenamefont {Andersen}\ \emph {et~al.}(2012)\citenamefont
  {Andersen}, \citenamefont {Esberg}, \citenamefont {Knudsen}, \citenamefont
  {Thomsen}, \citenamefont {Uggerh\o{}j}, \citenamefont {Sona}, \citenamefont
  {Mangiarotti}, \citenamefont {Ketel}, \citenamefont {Dizdar},\ and\
  \citenamefont {Ballestrero}}]{kandersen2012}%
  \BibitemOpen
  \bibfield  {author} {\bibinfo {author} {\bibfnamefont {K.~K.}\ \bibnamefont
  {Andersen}}, \bibinfo {author} {\bibfnamefont {J.}~\bibnamefont {Esberg}},
  \bibinfo {author} {\bibfnamefont {H.}~\bibnamefont {Knudsen}}, \bibinfo
  {author} {\bibfnamefont {H.~D.}\ \bibnamefont {Thomsen}}, \bibinfo {author}
  {\bibfnamefont {U.~I.}\ \bibnamefont {Uggerh\o{}j}}, \bibinfo {author}
  {\bibfnamefont {P.}~\bibnamefont {Sona}}, \bibinfo {author} {\bibfnamefont
  {A.}~\bibnamefont {Mangiarotti}}, \bibinfo {author} {\bibfnamefont {T.~J.}\
  \bibnamefont {Ketel}}, \bibinfo {author} {\bibfnamefont {A.}~\bibnamefont
  {Dizdar}}, \ and\ \bibinfo {author} {\bibfnamefont {S.}~\bibnamefont
  {Ballestrero}} (\bibinfo {collaboration} {CERN NA63}),\ }\href {\doibase
  10.1103/PhysRevD.86.072001} {\bibfield  {journal} {\bibinfo  {journal} {Phys.
  Rev. D}\ }\textbf {\bibinfo {volume} {86}},\ \bibinfo {pages} {072001}
  (\bibinfo {year} {2012})}\BibitemShut {NoStop}%
\bibitem [{\citenamefont {Xavier}(1990)}]{Artru1990}%
  \BibitemOpen
  \bibfield  {author} {\bibinfo {author} {\bibfnamefont {A.}~\bibnamefont
  {Xavier}},\ }\href {\doibase https://doi.org/10.1016/0168-583X(90)90122-B}
  {\bibfield  {journal} {\bibinfo  {journal} {Nuclear Instruments and Methods
  in Physics Research Section B: Beam Interactions with Materials and Atoms}\
  }\textbf {\bibinfo {volume} {48}},\ \bibinfo {pages} {278 } (\bibinfo {year}
  {1990})}\BibitemShut {NoStop}%
\bibitem [{\citenamefont {Nielsen}(2020)}]{CFN2019GPU}%
  \BibitemOpen
  \bibfield  {author} {\bibinfo {author} {\bibfnamefont {C.}~\bibnamefont
  {Nielsen}},\ }\href {\doibase https://doi.org/10.1016/j.cpc.2019.107128}
  {\bibfield  {journal} {\bibinfo  {journal} {Computer Physics Communications}\
  }\textbf {\bibinfo {volume} {252}},\ \bibinfo {pages} {107128} (\bibinfo
  {year} {2020})}\BibitemShut {NoStop}%
\end{thebibliography}
%merlin.mbs apsrev4-1.bst 2010-07-25 4.21a (PWD, AO, DPC) hacked
%Control: key (0)
%Control: author (8) initials jnrlst
%Control: editor formatted (1) identically to author
%Control: production of article title (-1) disabled
%Control: page (0) single
%Control: year (1) truncated
%Control: production of eprint (0) enabled
%

\end{document}